\documentclass[useAMS,usenatbib]{mn2e}

\usepackage{epsfig}
\usepackage{subfigure}
\usepackage{graphics}
\newcommand{\tdf}{2dFGRS}
\newcommand{\mn}{MNRAS}
\newcommand{\pasp}{PASP}
\newcommand{\apj}{ApJ}
\newcommand{\ha}{$\rm{H}{\alpha}$}
\newcommand{\hb}{$\rm{H}{\beta}$}
\newcommand{\hd}{$\rm{H}{\delta}$}
\newcommand{\rf}{${r}_{\rm{F}}$}
\newcommand{\bj}{${b}_{\rm{J}}$}
\newcommand{\rstarf}{${r}^*_{\rm{F}}$}
\newcommand{\msolar}{$\rm{M}_{\odot}$}
\title[Environments of 2dFGRS starbursts]{The environments and clustering properties of 2dFGRS-selected starburst galaxies}
\author[M.S. Owers et. al.]{Matt S. Owers,$^{1}$\thanks{E-mail: mowers@phys.unsw.edu.au}, Chris Blake$^{2}$, Warrick J. Couch$^{1,2}$, 
Michael B. Pracy$^{3}$, Kenji Bekki$^{1}$\\
$^{1}$School of Physics, University of New South Wales, Sydney, NSW 2052, Australia\\
$^2$Centre for Astrophysics and Supercomputing, Swinburne University of Technology, Hawthorn, VIC 3122, Australia\\
$^3$Research School of Astronomy and Astrophysics, Australian National University, Canberra, ACT 2600, Australia}
\begin{document}
\date{Accepted 2007 December 00. Received 2007 December 00; in original form 
2007 January 01}

\pagerange{\pageref{firstpage}--\pageref{lastpage}} \pubyear{2007}

\maketitle

\label{firstpage}

\begin{abstract}
We investigate the environments and clustering properties of starburst galaxies selected from the 2dF Galaxy Redshift Survey (2dFGRS) 
in order to determine which, if any, environmental factors play a role in triggering a starburst. We quantify the local environments, 
clustering properties and luminosity functions of our starburst galaxies and compare to random control samples. The starburst galaxies 
are also classified morphologically in terms of their broad Hubble type and evidence of tidal merger/interaction signatures. We find the
starburst galaxies to be much less clustered on large (5--15\,Mpc) scales compared to the overall 2dFGRS galaxy population. In terms of 
their environments, we find just over half of the starburst galaxies to reside in low to intermediate luminosity groups, and a further 
$\sim$30 per cent residing in the outskirts and infall regions of rich clusters. Their luminosity functions also differ significantly
from that of the overall 2dFGRS galaxy population, with the sense of the difference being critically dependent on the way their star 
formation rates are measured. In terms of pin-pointing what might trigger the starburst, it would appear that factors relating to their 
{\it local} environment are most germane. Specifically, we find clear evidence that the presence of a near neighbour of comparable 
luminosity/mass within 20\,kpc is likely to be important in triggering a starburst. We also find that a significant fraction (20--30 per 
cent) of the galaxies in our starburst samples have morphologies indicative of either an ongoing or recent tidal interaction and/or 
merger. These findings notwithstanding, there remain a significant portion of starburst galaxies where such local environmental influences
are not in any obvious way playing a triggering role, leading us to conclude that starbursts can also be internally driven.  
\end{abstract}


\section{Introduction}

Identifying and understanding the drivers of galaxy evolution remains one of the most important challenges in modern astrophysics. 
Whilst the model of hierarchical large scale structure and
galaxy formation provides a useful broad framework within which to understand how galaxies form and evolve, 
the impact of the environment on the evolution of a galaxy remains an important detail that is yet to be fully 
understood. One approach to gauging environmental influences on galaxy evolution is to obtain large samples of galaxies in the throes of
violent evolution enabling searches for statistically significant correlations between the galaxy properties and the 
environment it lives in. Thanks to large spectroscopic surveys such as the 2dF Galaxy Redshift Survey (2dFGRS; Colless et al. 2001) and 
the Sloan Digital Sky Survey (SDSS; York et al. 2000), such large samples are now accessible. 
Starburst galaxies (galaxies with star formation rates orders of magnitude above their quiescent rates) 
provide an excellent opportunity to study galaxies during rapid evolution, since the burst is generally expected to have 
been triggered within the last 500\,Myrs or so.

Starburst galaxies are found across the mass spectrum of galaxies, from the massive Ultra Luminous Infrared Galaxies (ULIRGs)
to dwarf starbursting galaxies such as the blue compact dwarfs (BCD) and dwarf irregulars (dIrr). There are a number of physical 
mechanisms proposed for triggering star formation across this broad mass spectrum, some of which are due to intrinsic processes, whilst 
others are due to environmental interaction. Spiral density waves are known to trigger star formation, however this mechanism is probably 
not strong enough to trigger a massive starburst in larger galaxies, whilst bars are effective at transporting gas to the central regions
of a galaxy triggering a circum-nuclear burst \citep{kennicutt98}. Whilst inadequate or absent in massive galaxies, internal 
processes such as stochastic self-propagating star formation \citep{gerola80} and cyclic gas re-processing \citep{papaderos96}
may cause dwarf galaxies to undergo multiple cycles of starbursts and quiescent phases on timescales that are short compared to a Hubble 
time. 

Major mergers of two or more gas-rich galaxies have been shown to produce massive bursts of star formation \citep{bekki01a,mihos96}, 
whilst more than $ 95\rm{\,per\, cent}$ of ultra-luminous infrared 
galaxies (ULIRGs) are merger systems (see \citet{sanders96} for a review). Numerous studies have been
conducted on the effects of close galaxy pairs on star formation rate (SFR), with the general consensus being pairs with
projected separations of less than $\sim 50$\,kpc show enhanced SFRs \citep{freedman06,nikolic04,lambas03}.
There is evidence that the mass ratio of the pairs may also be important, \citet{freedman06} find evidence that pairs with 
$\left|\Delta m_R \right|<2$ show strong correlations between SFR and projected separation, whilst those with 
$\left|\Delta m_R \right|\geq 2$ show no correlation, consistent with \citet{lambas03}. These results suggest that tidal interactions 
and mergers with galaxies of approximately equal mass are important in triggering star formation. 

The effects of the cluster environment may trigger starbursts; theoretical studies have shown that it is possible to trigger a 
starburst due to the high static thermal and ram pressure felt by molecular clouds within a galaxy disc due to the tenuous, hot 
intra-cluster medium (ICM) \citep{bekki03}, whilst group member galaxies may experience starbursts during the infall process 
due to the time dependant tidal gravitational field \citep{bekki99}. The above mechanisms can be tested observationally, since the 
different triggering mechanisms are expected to produce different spatial distributions of the starburst across the galaxies. Much 
observational work has been carried out on the fractions of the types of galaxies residing in clusters, and on the populations in major 
cluster mergers. Photometric studies revealed significant evolution of the blue population with redshift (the Butcher-Oemler effect, 
\citet{butcheroemler}). Subsequent spectroscopic studies revealed the effect was mostly caused by galaxies which had undergone a burst 
of star formation between 0.1 and 1.5\,Gyrs prior to the epoch of observation \citep{couch87}. 

At the low mass end of starburst galaxies are the BCDs where much work has been carried out in order to 
determine triggering mechanisms. \citet{pustilnik01} find that a significant fraction ($\sim80\rm{\,per\, cent}$) of a sample of 86 BCDs 
either have neighbours close enough to induce strong tidal forces, or display a disturbed merger morphology, implying that interactions
are important in triggering starbursts in BCDs. \citet{noeske01} draw a similar conclusion from a sample of 98 star forming 
dwarf galaxies, finding a lower limit of $\sim 30\rm{\,per\, cent}$ harbour close neighbours, whilst \citet{taylor97} find the rate of 
companion occurrence in HII galaxies is $\sim 57\rm{\,per\, cent}$, more than twice the rate of companions in a comparison sample of low 
surface brightness galaxies. However, \citet{brosch04} find that in a sample of 96 BCG and dIrrs imaged in \ha\, there is not a 
significant number of star forming companions, concluding that tidal triggering is probably not important in enhancing star formation in 
these galaxies. Based on the observation that the star formation seems to be occurring at the outer boundaries of the dIrrs,
\citet{brosch04} propose a `sloshing' mechanism for triggering the star formation, where gas oscillates within the dark matter halo, 
producing gas build up at locations where the gas `turns around' providing a trigger for star formation. In any case, the triggering 
mechanism for these dwarfs is still under much debate.

Also of relevance in this context is the population of post-starburst galaxies, first discovered in distant clusters and called 
`E+A' galaxies by Dressler \& Gunn (1982, 1983). These galaxies are known as post-starburst galaxies, since their rather enigmatic 
spectra -- characterized by a strong A-type stellar spectrum superimposed upon a normal E galaxy type spectrum and devoid of any 
emission lines -- would indicate they have undergone a recent burst of star formation that was abruptly truncated within the last 
1\,Gyr (Couch \& Sharples 1987, Poggianti et al. 1999). Subsequent studies have shown that E+A galaxies are also found in a variety of 
environments at lower redshifts (e.g. Zabludoff et al. 1996), and of particular note is the 2dFGRS-based study by \citet{blake}. They 
identified all the E+A galaxies within the \tdf\ and studied their local and global clustering environments in an attempt to pin-point 
what might have triggered the original starburst. They concluded that the environments of E+A galaxies do not differ significantly from 
that of the general 2dFGRS galaxy population as a whole. However, the morphologies are consistent with a merger origin, and hence the 
triggering/cessation mechanism for the initial burst is likely to be driven by very local encounters. Simulations have shown that the 
E+A signature marks the late stage of a galaxy-galaxy merger \citep{bekki01b} where the merging companions are no longer separable. The 
next logical step, therefore, is to study the environments of the starburst galaxies, which are thought to be the progenitors of E+A 
galaxies. 

Motivated by this need and the broader issues outlined above as to what mechanisms are responsible for triggering starburst events, this 
paper reports a systematic study of the environments and clustering properties of the starburst galaxies within the 2dFGRS, that very 
much parallels the E+A study by \citet{blake}. We select two starburst samples, using complementary star formation rate determinants, 
with each sample containing 418 galaxies. This allows measurement of statistical properties of the starburst population, such as the 
luminosity function, near neighbour properties, morphological distribution and large scale clustering. The paper is organised as follows: 
In Section \ref{selection} we briefly introduce the \tdf\, and describe the selection criteria for our starburst samples. In Section 
\ref{morphologies} we use the Supercosmos Sky Survey (hereafter SSS) images to study the morphologies of our samples. In Section 
\ref{lumfun} we measure the \bj\, band luminosity function. In Section \ref{localenv} we investigate the small scale environment
surrounding the starbursts. In Section \ref{largescale} we measure large scale environmental properties of the sample. In Section 
\ref{discussion} we summarise and discuss our findings. Throughout this paper, we assume a standard $\Lambda \rm{CDM}$ cosmology where 
$\Omega_m=0.3, \Omega_{\Lambda}=0.7$ and $h=H_0/(100\,\rm{km\,s}^{-1}\,\rm{Mpc}^{-1})=0.7$. We convert redshifts to physical distances 
using these parameters.

\section{Sample selection}
\label{selection}
\subsection{The 2dF Galaxy Redshift Survey}
Our starburst samples were selected from the final release of the \tdf, a large spectroscopic survey conducted at the 
Anglo-Australian Telescope using the Two-Degree Field (2dF) multi-fibre 
spectrograph. The 2dF instrument is capable of simultaneously observing 400 
objects over a $2^{\circ}$ diameter field of view. The survey consists of 
$\sim$220,000 galaxies covering an area of approximately 2,000 deg$^2$ in 
three regions: a South Galactic Pole (SGP) strip, a North Galactic Pole (NGP) 
strip and a series of 99 random fields dispersed around the SGP. The input 
source catalogue for the 2dFGRS was derived from an extended version of the 
Automated Plate Measuring \citep[APM;][]{maddox90a,maddox96} galaxy catalogue based on APM machine scans 
of 390 plates from the UK Schmidt Telescope (UKST) Southern Sky Survey. The 
extinction corrected survey limit was \bj=19.45 and the 
median redshift is $\overline{\rm z}=0.11$. The spectra are collected through 
2\,arcsec fibres, cover the wavelength range $3600-8000$\AA, have two-pixel 
resolution of 9.0 \AA\ and a median $S/N$ of 13 $\rm{pixel}^{-1}$. The large 
range in wavelength is made possible by the use of an Atmospheric Dispersion 
Compensator (ADC) within the 2dF instrument. Redshift determinations are 
checked visually and allocated a quality parameter, Q, ranging from 1 to 5. 
Redshifts with Q $\geq3$ are regarded as reliable at the 98.4\rm{\,per\, cent} level. The rms 
uncertainty of the redshifts is $85\,\rm{km}\ \rm{s}^{-1}$. The spectra are not 
flux calibrated and consist of a sequence of `counts' as a function of 
wavelength. The 2dFGRS data base can be accessed at http://www.mso.anu.edu.au/2dFGRS.

\subsection{The \tdf\, spectral line catalogue}
We used the \tdf\, spectral line catalogue prepared by Ian Lewis to select our 
samples. The fitting procedure is described in detail by \citet{lew1}. Here we briefly
outline the fitting procedure as follows: The continuum in the vicinity of each spectral line was removed by subtracting 
the median level over a $133$\,\AA\, window. Gaussian profiles were fitted for up to 20 lines per spectrum in both emission
and absorption, with the height, width and a small perturbation in redshift (always $\Delta\rm z < 0.005$) as free parameters. 
After fitting, the quality of the fit was determined using the parameters obtained from the 
fit (height, width and area) and the rms residuals. A quality flag was assigned
to each line fit in the range 0 (bad fit) to 5 (good fit) and a signal to noise 
parameter was calculated for each line by taking the average signal per pixel and dividing by the average noise per pixel for pixels 
lying within $\pm3\sigma$ of the line center. Equivalent 
widths (EWs) were calculated using the measured flux in the line and the fitted
continuum level (in the $133$\,\AA\, window local to the line). The EWs can be used 
in line ratio diagrams without the need for flux calibration. However, it 
is assumed there is no significant flux contribution to the continuum level from effects 
such as scattered light. We correct for broadening of the spectral lines due 
to redshift by dividing the EWs by 1+z, where z is the galaxy 
redshift.
\subsection{Selection of emission line galaxies}
In order to select a high quality sample of starburst galaxies, we first 
ensured the parent catalogue consisted of single 
observations of galaxies with high quality spectra. Thus, we kept 
only spectra with:
\begin{itemize}
\item ADC=1\footnote{Prior to August 1999, the ADC suffered positioning problems resulting in the blue end of each spectrum being
severely depleted of counts. All spectra obtained during this period have been assigned ADC flag =0}
\item \tdf\, redshift parameter,Q $\ge3$
\item $\rm{z} \ge 0.002$ to exclude stars.
\end{itemize}
The parent \tdf\, spectral line catalogue contains 264,765 spectra, and the 
exclusion of those objects failing the above criteria reduced this to 172,427 spectra. 
We then included only the highest quality spectra of the 
multiply observed objects based on the redshift quality flag or the highest 
mean S/N measured between 4000 and 7500\,\AA\, (if the redshift quality flags were equal), leaving a parent catalogue of 
162,529 high quality spectra with no multiple observations.

From this catalogue, we selected a sample of emission line galaxies from which 
a sub-sample of star forming galaxies was selected.
It is standard practice to differentiate star forming galaxies from  
narrow-emission-line AGNs according to their position on the BPT diagram 
\citep{bpt}, where the logarithm of the ratio's of [OIII]/\hb\, and 
[NII]/\ha\, are plotted. With this in mind, we selected galaxies with EW$\geq2.5$\,\AA\, in 
emission, line S/N$>1.0$ and a line fit quality flag $\geq 4$ for the following species: \ha,  NII, OIII and \hb. The cut 
of EW$\geq2.5$\,\AA\, is arbitrary, but ensures a robust line detection for use in the
line ratio diagram. We detected 9,516 emission line galaxies that conform to the 
above criteria; Figure 1 shows them plotted in the BPT diagram. The advantage of the chosen lines for the ratios plotted
is three-fold. Firstly, the lines are close together in wavelength space, hence extinction
due to dust has little effect on the calculated ratios. Secondly, the change in the continuum level is minimal for nearby lines, meaning
the ratio is essentially equivalent to a line flux ratio. Thirdly, as seen in Figure 1 the
plot of the ratios show two distinct populations in the [OIII]/\hb\,-- [NII]/\ha\, plane, the
origin of which lies in the different physical mechanisms providing the photo-ionizing continuum.
For AGNs, the ionizing radiation is non-thermal in nature and is well approximated by 
a power law, whilst the ionizing radiation in star-forming galaxies is in the form of UV radiation
emitted by hot OB type stars (see \citet{veil} for a detailed description). The empirically 
derived starburst/AGN classification line of Kauffmann et al. (2003) was used to select a sample of galaxies 
which have emission dominated by star-formation (although we note there may be a small contribution to emission line strength due 
to AGN emission) such that a star forming galaxy is one which has
\begin{equation}
\rm{Log}(\frac{[OIII]}{H_{\beta}}) < \frac{0.61}{\rm{Log}([NII]/H_{\alpha})-0.05}+1.3.
\end{equation}
 This sample contains 8,369 galaxies from which our starburst samples are selected. 
\begin{figure}
\center
\epsfig{file=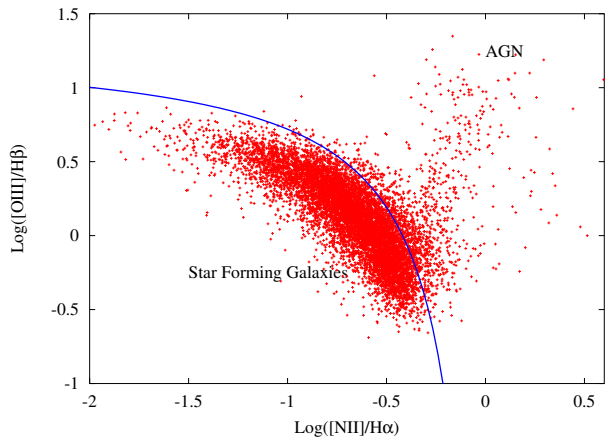,height=\linewidth,angle=-90}
\caption{BPT emission line ratio diagnostic diagram, with all the 2dFGRS emission line galaxies plotted as {\it dots}. The {\it solid 
line} represents the empirically derived star formation/AGN demarcation line. Due to the non-thermal mechanism for the underlying 
photo-ionising continuum, AGNs inhabit a different region of this diagram compared to star-forming galaxies, as can be seen by the plume 
of galaxies to the {\it upper right}. Galaxies residing in this plume are excised from our sample as their emission is not due to star 
formation.}
\label{BPTfig}
\end{figure}

It is pertinent to note here that the \tdf\, fibres cover only a 2--2.16\,arcsec angular diameter on the galaxies (4\,kpc at the median 
\tdf\, redshift of $\overline{\rm z}=0.11$), and our sample will not contain galaxies where star formation is not occuring within the 
area covered by the fibre aperture. However, the mean RMS error in the fibre positioning was 0.3\,arcsec, whilst the seeing was of the
order 1.5--1.8\,arcsec, hence the observed spectra cover a larger fraction of the galaxy once these effects are accounted for, and the 
aperture effect is somewhat diluted. We also note that since we do not obtain the total \ha\, flux of each galaxy, accurate total star 
formation rates cannot be derived. \citet{kewley05} showed that properties such as total aperture corrected SFRs derived from 
\tdf\, spectra would be most affected at $z<0.06$, whilst \citet{hopkins} found that those galaxies with the largest aperture corrections
(ie. those with the largest angular size) had their \ha\, SFRs slightly overestimated when compared to the 1.4\,GHz SFRs. Thus, we expect 
galaxies at lower redshifts will have their aperture corrected \ha\, SFRs slightly overestimated. In section \ref{sfrtot} we attempt to 
correct for this and other redshift dependencies by binning the data in redshift space when selecting our starburst sample. The
combination of redshift binning and the selection of only the most extreme galaxies in SFR per redshift bin should serve to minimize the 
aperture effect, although we note there may be some contamination in our sample.

\begin{figure}
\center
\epsfig{file=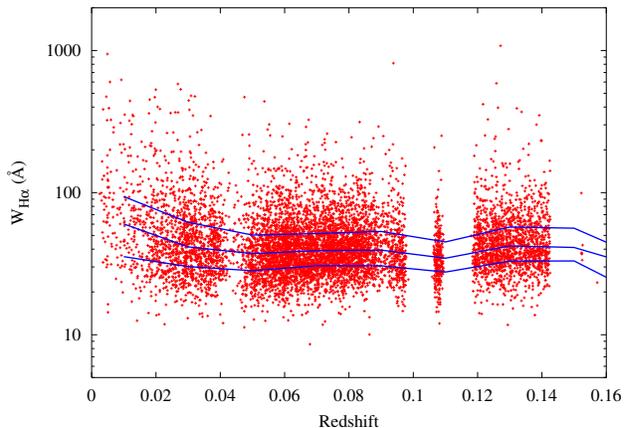,height=\linewidth,angle=-90}
\caption{Plot of $W_{H\alpha}$ as a function of redshift. The {\it solid lines} show (from bottom to top) the first, second (median) and 
third quartiles. They show that there is no appreciable trend with redshift, apart from at low redshifts where the turn-up in 
$W_{H\alpha}$ is possibly due to contamination by extra-galactic HII regions and dwarf starbursting galaxies that are not detected at 
higher redshifts. The gaps in the distribution are due to redshifting of emission lines used in the selection process into areas
where sky dominates the spectrum. This causes the lines to have bad fits, and thus be rejected by our criteria.}
\label{ew.vs.z}
\end{figure}

Figure \ref{ew.vs.z} shows that the \ha\, EW (hereafter $W_{H\alpha}$) does not show any trends with redshift in the range 
$0.04<z<0.16$. However, we find  
evidence of a slight increase in the lowest redshift bin ($0.002<z<0.04$) with the median $W_{H\alpha}$ increasing from ~40 to ~60. This 
increase is due to low mass emission line systems such as HII regions and dwarf starbursting galaxies which, due to their low luminosity,
would not be detected at higher redshifts.  Interestingly, \citet{mateus04} find no trends in $W_{H\alpha}$ with redshift 
out to z=0.05. The gaps in the $W_{H\alpha}$ distribution correspond to redshifts where emission lines used 
in the selection criteria are shifted into areas where sky emission/absorption lines dominate the spectra, producing bad line fits meaning 
these galaxies are rejected during the selection process. $W_{H \alpha}$ is a relative quantity meaning the effect of aperture depends 
on both the spatial distribution of the \ha\, line strength and the continuum profile in the galaxy. 
We assume this distribution is uniform and we make no attempt to correct for aperture in the starburst sample in Section \ref{normsfr}. 
We note there will be contamination of our sample due to HII regions, however this contamination is negligible and makes up only 1.6 per 
cent of the sample selected in Section \ref{normsfr}.

\subsection{Starburst galaxy selection criteria}

Starburst galaxies are generally defined by their extremely high SFRs, such that if the SFR was sustained,
the gas reservoirs feeding the burst will be exhausted in timescales much shorter than the Hubble time, typically of the order of
$10^8$\,yrs. The problem then becomes selecting observable quantities that adequately fit the above definition. In general, an
observable quantity is converted into an SFR via some theoretical model which makes various assumptions. Some examples of these 
observables are: 
\ha\, luminosity, infrared luminosity, radio emission, a galaxies position on the [OII]--\hd\, plane, [OII] luminosity, UV luminosity etc 
(see eg. \citet{sullivan,kennicutt98,hopkins}). A galaxy is then generally defined as a starburst based on some arbitrary cutoff in 
absolute SFR, SFR per unit area or the Scalo b-parameter (ie. the ratio of current to average past SFR). Below we describe two samples 
selected from the emission line galaxies above using one approach equivalent to measuring the b-parameter and one equivalent to measuring
the absolute SFR.

\subsubsection{Normalized SFR sample}
\label{normsfr}
The first method employed to estimate the SFR of the emission line galaxies selected above
uses the method of \citet{lew1} to determine a normalized SFR, $\mu^*$, defined by
\begin{equation}
\mu^*=\eta W_{H\alpha} L^*,
\end{equation}
where $\eta = 8.2\times10^{-42}$ \msolar\,$\rm{s}\, \rm{ yr}^{-1}\,\rm{erg}^{-1}$ is the conversion 
from $L_{H\alpha}$ to SFR \citep{sullivan}, 
$L^*=1.1\times10^{40} \rm {erg}\, \rm {s}^{-1}\, \rm {\AA}^{-1}$ is the luminosity corresponding to 
the knee in the $r'$ band luminosity function as determined by \citet{blanton} and $W_{H\alpha}$ is the 
equivalent width of the \ha\, emission line ($W_{H\alpha}\approx L_{H\alpha}/L_{c}$, 
$L_{c}$ is the continuum luminosity). The SFR measured in this way gives a SFR normalized
to $L^*$. A 1\AA\, correction is made to $W_{H\alpha}$ to account for stellar absorption \citep{hopkins}. Thus the 
SFR normalized to $L^*$ is given by $\mu^*=0.0902(W_{H\alpha}+1)$. 

\begin{figure*}
\center
\includegraphics[height=\textwidth,angle=-90]{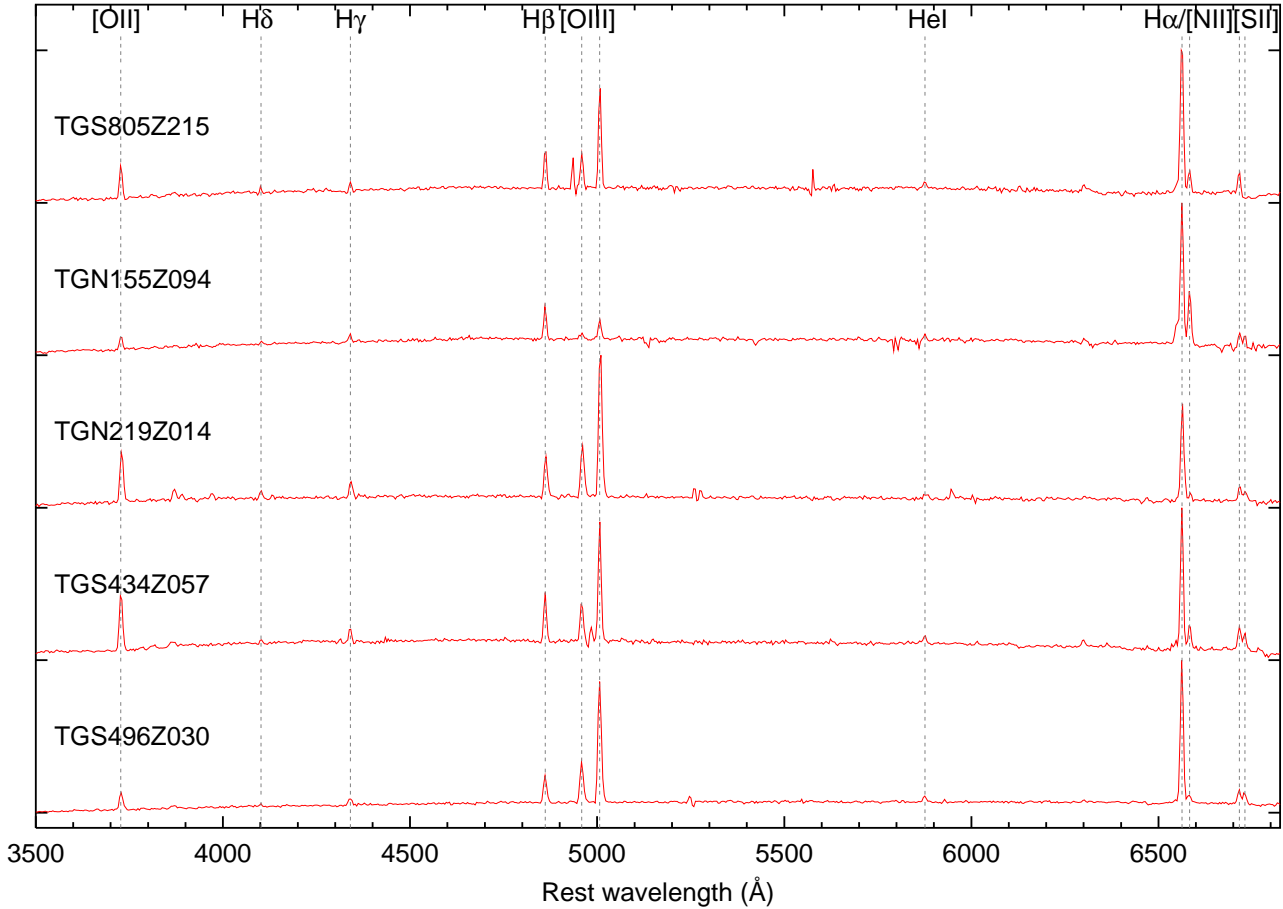}
\caption{Example spectra for starburst galaxies within our SFRNORM sample (see Fig. \ref{morphs} for corresponding SSS images). The 
wavelength scale is in the galaxy rest frame. Prominent night-sky emission lines or absorption features have been excised and 
interpolated over. {\it Dashed vertical Lines} show the positions of prominent spectral features.}
\label{normspec}
\end{figure*}

It is not clear where the dividing line between a normal star forming galaxy and a starbursting galaxy can be drawn using this method for
determining the SFR, hence we arbitrarily defined our starburst sample to be the top 5 per cent of galaxies in terms of their $\mu^*$ 
values; hereafter we refer to this as our SFRNORM sample. This corresponds to a cut at 
$\mu^*=9.6$\,\msolar\,$\rm{ yr}^{-1}$, or a $W_{H\alpha}$ of $\sim 106$\,\AA. $W_{H\alpha}$ is essentially a measure
of the ratio of the SFR to the mass of the old stellar population, and it
correlates strongly with the scalo b-parameter \citep{ken94}. The lower limit on $b$ corresponding to our $W_{H\alpha}$ is  
$\sim 2$ (using Table 1 of \citet{ken94} with a Salpeter IMF). Since quiescent disks have $b\sim0.5$, this seems a reasonable 
cutoff value to delineate starbursts \citep{kennicutt05}. Figure \ref{normspec} shows example spectra selected from the SFRNORM sample 
where the spectra are de-redshifted and normalized such that the strongest emission line peaks at 1. Dominant emission species are
marked on the plots.

\subsubsection{Total SFR sample}
\label{sfrtot}
It is common to flux calibrate spectra using broad band photometry. We used
\rf\, magnitudes obtained from the SSS to estimate the continuum level at the 
wavelength of \ha. The \rf\, magnitudes were K-corrected to their rest frame values using 
equation 37 of \citet{wild} and the K-corrected values were converted to $r_{AB}$ magnitudes with 
equations A12 and A16 of \citet{cross}. We use equation 15 of \citet{blanton} to derive
the continuum luminosity, 
\begin{equation}
L_{cont}=4.18\times 10^{24} ({\rm {W\,\AA^{-1}}})\times10^{-0.4M_{AB}},
\end{equation}
where $M_{AB}$ is the absolute AB magnitude and the corresponding effective wavelength
for the $r_{AB}$ band is $\lambda_{eff}=5595$\,\AA. The \ha\, luminosity is then derived from $L_{cont}$
and the \ha\, equivalent width such that 
\begin{equation}
L_{{H\alpha}}=2.5L_{cont}\times({W}_{H\alpha}+1), 
\end{equation}
where ($W_{H\alpha}$) is corrected by 1\,\AA\, to allow for the negative contribution of stellar absorption \citep{hopkins} and the 
factor of 2.5 corrects for $\sim 1$ magnitude extinction of the $r_{AB}$ magnitude 
due to dust extinction, consistent with the extinction measured at \ha\, by \citet{ken83} and \citet{Niklas}. The star formation rate 
(SFR) is related to $L_{H\alpha}$ by eqn. 1 of \citet{sullivan};
\begin{equation}
SFR({\rm M_{\odot}\,yr^{-1}})=8.2\times10^{-35}L_{{H\alpha}}{\rm(W)}. 
\end{equation}
The discrepancy between the effective wavelength
of the $r_{AB}$ band and the wavelength of the \ha\, emission should not have a large effect on our
measured SFRs since the gradient of the continuum at these wavelengths is close to zero. This 
method effectively corrects for aperture effects, assuming the $W_{H\alpha}$ is constant
across the galaxy (see \citet{hopkins}), hence it is equivalent to measuring the total absolute SFR, hereafter referred to as SFRTOT.

\begin{figure*}
\center
\includegraphics[height=\linewidth,angle=-90]{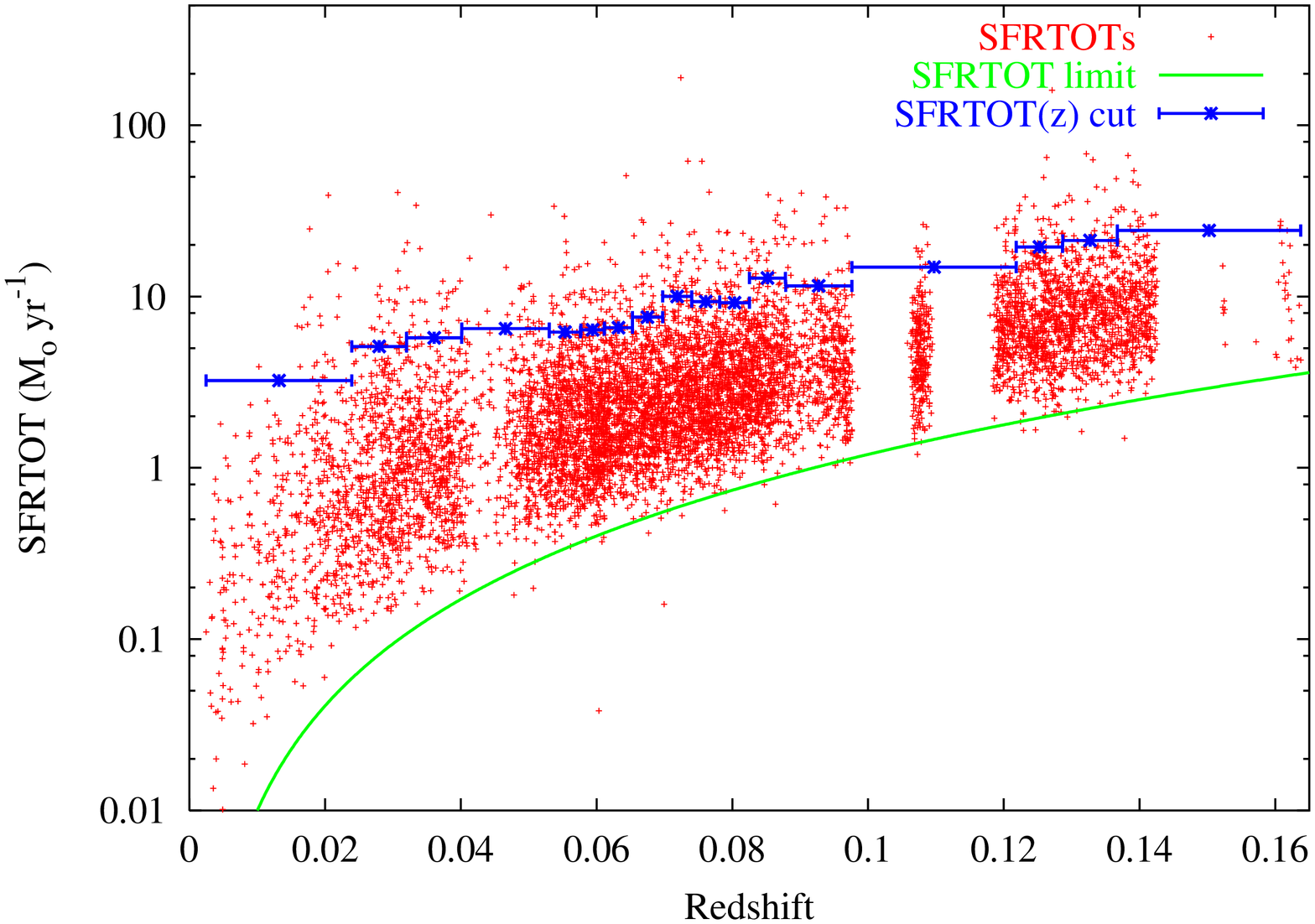}
\caption{Plot of the total star formation rates (SFRTOT) measured for the 2dFGRS galaxies versus redshift. The {\it green} line indicates
the effective SFR limit derived from the apparent \rf\, band magnitude limit and $W_{H\alpha}\sim 20$ corresponding to
approximately the lowest measured $W_{H\alpha}$ in the star forming sample. The {\it blue} horizontal bars indicate our binning of the 
data in redshift (to ensure equal statistical weight to each bin) with their vertical height indicating the top 5 per cent cutoff for defining 
our SFRTOT starburst sample.}
\label{sfrtotfig}
\end{figure*}
\begin{figure*}
\center
\includegraphics[height=\linewidth,angle=-90]{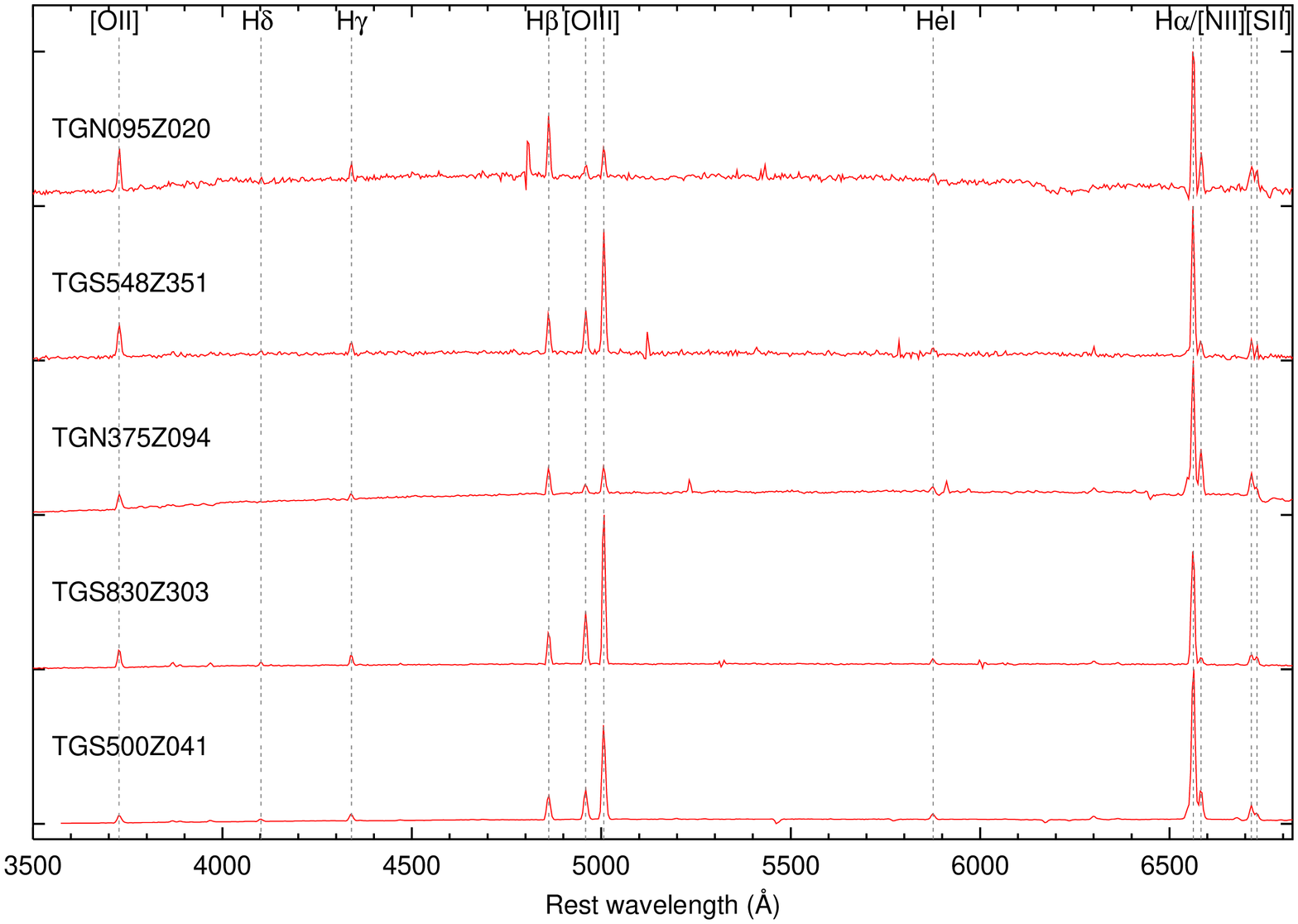}
\caption{Example spectra of starburst galaxies in our SFRTOT sample, displayed in the same manner as Fig. \ref{normspec}.}
\label{sfrtotspec}
\end{figure*}

Figure \ref{sfrtotfig} shows the dependence of the SFRTOT values on redshift. There exists a strong correlation with redshift, part of 
which can be explained by a combination of the survey magnitude limit, and a lower cut-off in $W_{H\alpha}$ at around 20\,\AA. This
produces an effective limit in the SFR observed with redshift. It is also known that below $z \sim 0.06$, aperture effects cause 
discrepant measures of SFR using this method \citep{kewley05,hopkins}. We attempt to nullify any redshift dependency by binning
the sample in redshift space, such that each redshift bin contains 500 galaxies (to ensure equal statistical weight), and selecting
the galaxies in the top $5\rm{\,per\, cent}$ of the SFRTOT distribution within each bin. This serves to select the most extreme 
star-forming galaxies in each redshift bin, and hence defines our second (`SFRTOT') starburst sample. Figure \ref{sfrtotfig} shows the 
cutoff in SFRTOT for each bin, where again the gaps correspond to regions where emission lines are redshifted into sky dominated portions
of the spectrum, hence are rejected by our selection criteria due to bad line fits. Example spectra of a sub-sample of galaxies selected 
from the SFRTOT sample are presented in Figure \ref{sfrtotspec} in a similar fashion to Figure \ref{normspec}. As for the SFRNORM sample, 
these spectra are clearly dominated by emission lines produced both by photo-ionization and forbidden transitions.
\subsubsection{Pristine SFRTOT starburst sample}
The above samples select out starburst galaxies based on the SFR derived from \ha\, alone. The advantage here is that \ha\, is less 
affected by dust than bluer lines such as \hd\, and [OII] -- both common tracers of recent star formation. Thus in the above, there is no 
restriction placed on the \hd\, and [OII] lines. This means we are not biased against selecting `dusty' starbursts which have 
their emission at bluer wavelengths heavily attenuated. However, this also means we may be selecting out older starburst 
galaxies which have been forming stars at a high rate for a several hundred million years. Indeed, it was noticed that a high proportion
of the SFRTOT sample had \hd\, in absorption. This may mean we are selecting a combination of both dusty starbursts and older 
starbursting systems. To test for any effect this may have, we have selected a sub-sample from the SFRTOT sample with the criteria that 
\hd\, is either in emission ( EW$\geq0$ and quality flag $\geq4$) or is not detected in absorption or emission (its quality flag is 0), 
and hence the nebular emission is dominating the stellar absorption. This ensures we are selecting a sample where the majority of the 
light emitted comes from a starburst that started $\sim 200$ Myrs ago \citep{balogh99}, hence we are more likely to find evidence of a 
triggering mechanism. This added selection criteria produces a sub-sample of 244 galaxies within the SFRTOT sample, which we will call 
the SFRTOT+ sample. In the following analysis, we present results for the extra sub-sample in addition to the main SFRTOT sample. A 
sub-sample of the SFRNORM sample using the above criteria will not be presented, since $97\rm{\,per\, cent}$ of the SFRNORM galaxies meet
the above criteria.

\section{Morphologies}\label{morphologies}

Of relevance to our study is the morphology of our different samples of starburst galaxies and, in particular, whether there is direct 
evidence that mergers and tidal interactions are responsible for driving the starburst activity. All the galaxies in our SFRNORM and 
SFRTOT starburst samples were therefore morphologically classified by one of us (W.J.C.) by visual inspection of blue (\bj) and red (\rf)
images obtained from the SSS. The SSS has digitised sky survey plates taken with the UK Schmidt telescope, with 
a pixel size of 0.67\,arcsec. The data were accessed using the website http://www-wfau.roe.ac.uk/sss/pixle.html. A description of the SSS 
is given by Hambly et al. (2001).

\begin{figure*}
  \begin{center}
    \begin{tabular}{ll}
      {\includegraphics[angle=0,width=0.48\linewidth]{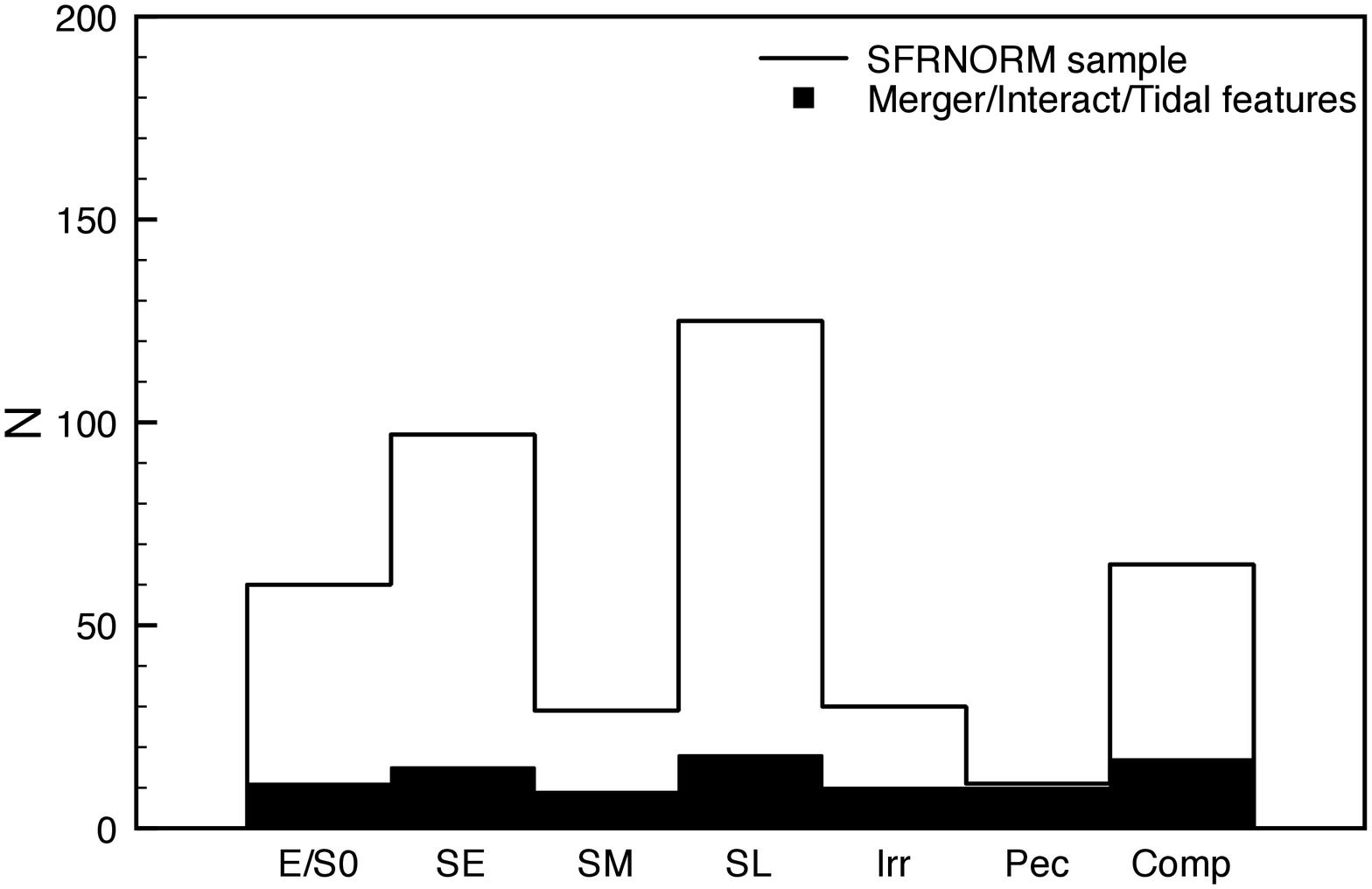}}&
      {\includegraphics[angle=0,width=0.48\linewidth]{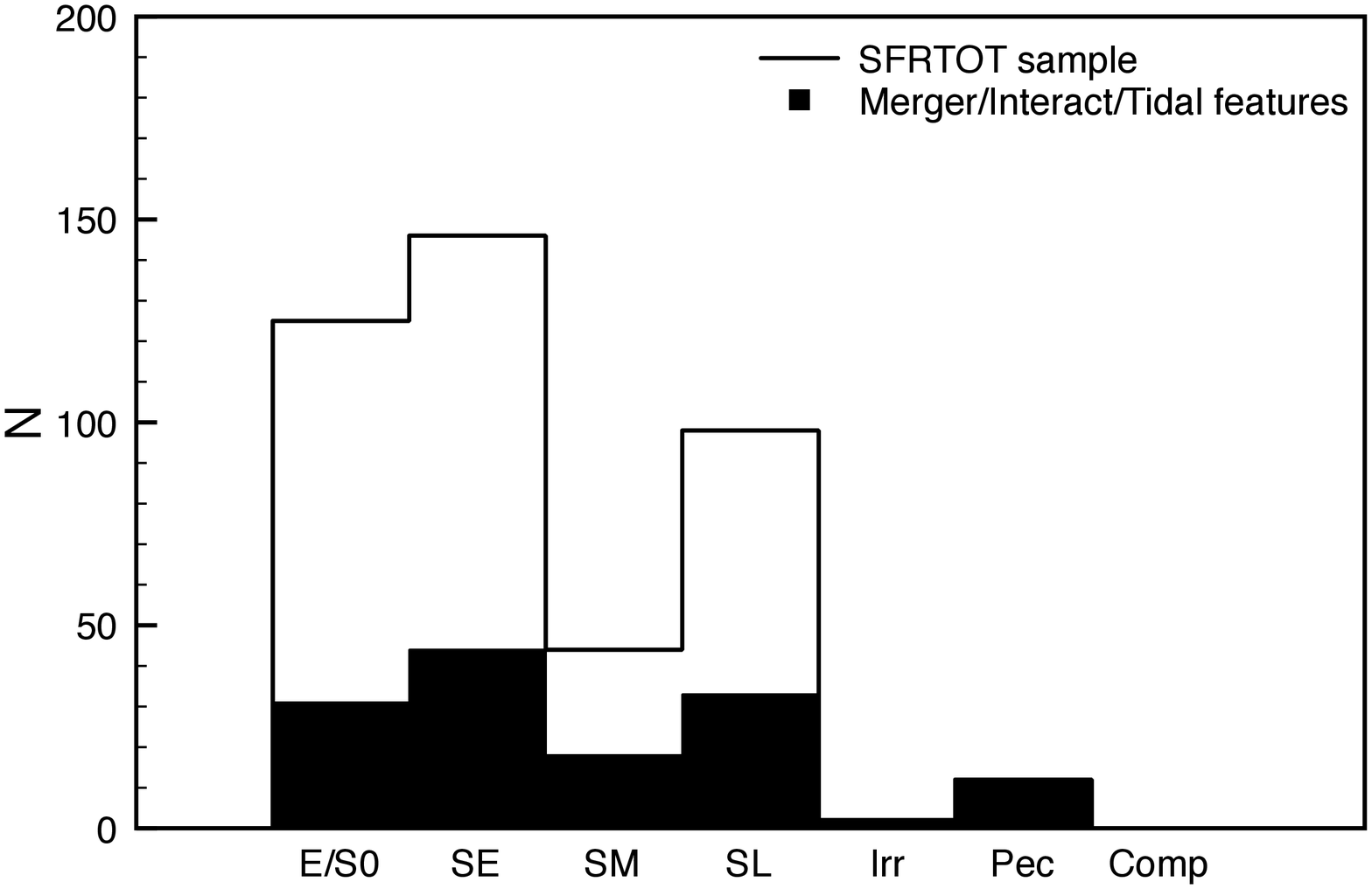}} \\
     \end{tabular}
    \caption{Histograms of the number of galaxies per morphological type within the SFRNORM ({\it left} panel) and SFRTOT ({\it right} 
panel) samples. The {\it filled} portions of the histograms show the number of galaxies within that particular class which showed 
evidence of a merger or some form of tidal or interaction activity.}
    \label{morph.hist}
  \end{center}
\end{figure*}

\begin{figure*}
\center
\includegraphics[width=\linewidth]{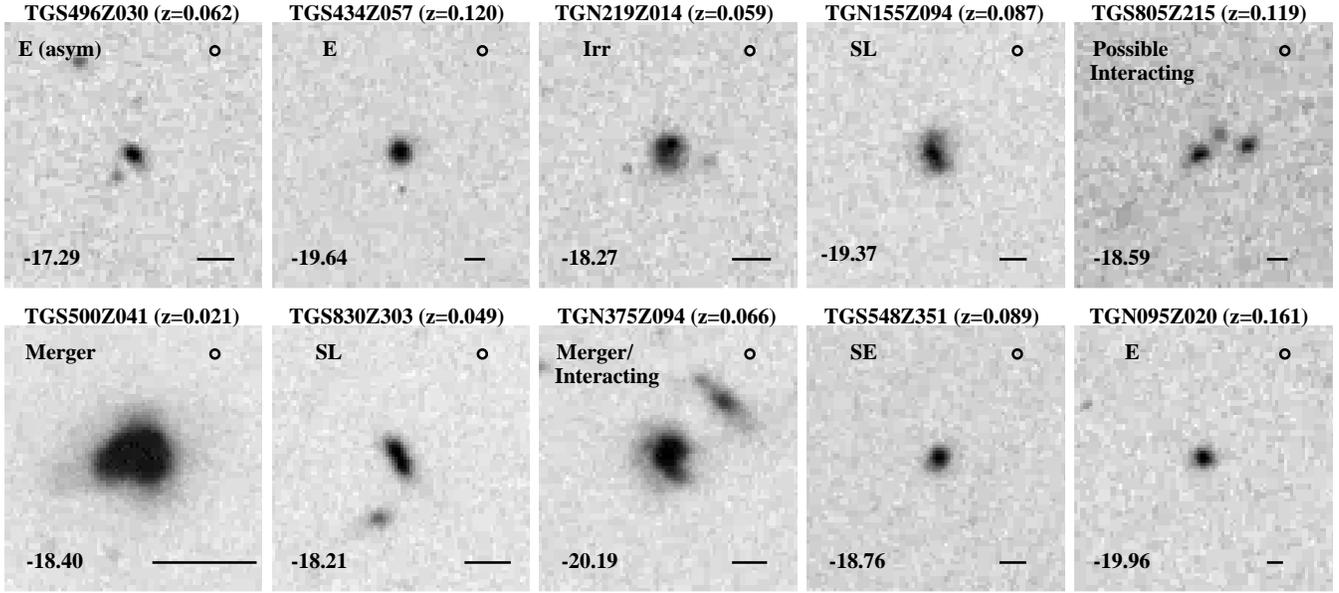}
\caption{SSS \bj\, band 1x1 $\rm{arcmin}^{2}$ images of the starburst galaxies whose spectra are shown in Figs. 3 and 5. The {\it top row}
contains the images for the SFRNORM galaxies shown in Fig. 3 while the {\it bottom row} contains the images for the SFRTOT galaxies shown 
in Fig. 5. The {\it bar} on the lower right of each image corresponds to 10\,kpc at the galaxy redshift. The {\it circle} shown at the 
top right hand corner indicates the size of the 2\,arcsec 2dF fibre. The absolute \bj\, magnitudes are given in the bottom left corner 
(as $M_b-5 {\rm{log}}_{10}h$). The morphological classifications from Section \ref{morphologies} are given in the top left corner}
\label{morphs}
\end{figure*}

The scheme used to morphologically classify our starburst samples was as follows: The first step was to try and assign a basic Hubble 
type to each galaxy. In doing so, the following five broad categories were used: E or S0 (E/S0), early-type spiral (SE), medium-type 
spiral (SM), late-type spiral (SL) and irregular (Irr). Using these broader categories, rather than attempting to use all the individual 
Hubble types (Sa, Sab, Sb, etc), was considered to be sufficient for our purposes and really all that was practical given the limited 
quality and resolution of the SSS images. The numbers of galaxies classified into each of these five morphological bins are plotted in 
histogram form in Figure \ref{morph.hist} and listed as percentages in Table 1, for both the SFRNORM and SFRTOT samples. We see that both
samples have a broad mix of morphologies, with no one type being dominant. Only in the SFRTOT sample do we see the absence of a 
particular morphological class, which somewhat surprisingly is the Irr types. Figure \ref{morphs} shows example \bj\, band SSS images of 5 
galaxies each from the SFRNORM and SFRTOT samples (corresponding spectra are presented in Figures \ref{normspec} and \ref{sfrtotspec}). 
The plots are labelled with the morphological classification, absolute \bj\, magnitude, redshift and \tdf\, label. No morphological
selection has been applied in selecting the starbursts appearing in this figure.

Whilst the majority of galaxies in both samples could be assigned a Hubble type, there were nonetheless a subset of galaxies where this 
was not possible. Such galaxies fell into one of two cases: (i) They were sufficiently well resolved to discern their structure, but that
structure was so peculiar that it did not resemble any of the Hubble types. These cases are denoted `Pec' in Figure \ref{morph.hist} 
and Table 1, where it can be seen that they represented only $2-3\rm{\,per\, cent}$ of each sample. (ii) They were too small to be 
resolved and hence be classified. Notably, we only encountered galaxies of this type -- which we denote as `Comp' (for compact) -- in 
the SFRNORM sample, where they account for $16\rm{\,per\, cent}$ of these types of starburst galaxies. The `Comp' classified galaxies 
can be separated into two groups: (i) The low redshift ($z<0.05$) galaxies whose compact nature, conspicuously blue colours (as seen on 
the SSS images), and their generally faint, `dwarf'-like luminosities (${M_{{b}_{\rm{J}}}}>-18$), suggest they are most likely blue 
compact dwarf (BCD) galaxies \citep{thuan81}. (ii) Those galaxies with $z>0.05$ and ${M_{{b}_{\rm{J}}}}<-18$ where the SSS images are 
simply not of high enough quality to assign a Hubble type.

The second step of our morphological classification process was to assess the images for evidence of on-going or recent merger 
and/or tidal interaction activity. To be classified as an on-going merger, a galaxy had to exhibit clear visual evidence of it coalescing
with  another, as manifested by the presence of tidal bridges and/or arms. Similarly, to be classified as being involved in a tidal 
interaction, a galaxy had to show clear signs of a tidal link (via a bridge or arm) with another galaxy. As such, we mainly identified 
what would be called `major' mergers/interactions (where the two galaxies involved have similar luminosities), given that it is only 
these types that are sufficiently conspicuous to satisfy the rather conservative visibility criterion that we took. Finally, evidence of 
a recent merger or interaction was inferred from the presence of tidal tails, shells or debris. This was by far the most difficult and 
subjective classification to make. Indeed, sometimes the evidence for a merger or interaction, while being  highly suggestive, was not 
always totally convincing, and such cases were placed into a second category called `Possible Merger/Interaction'. 

The numbers of starburst galaxies classified as mergers/interactions in this way are shown by the dark shaded histograms in Figure 
\ref{morph.hist} and listed as percentages in the last two columns of Table 1. Importantly, we find that a significant fraction of our 
starburst galaxies are involved in on-going or recent mergers and interactions. The highest number is found in the SFRTOT sample where 
$28\rm{\,per\, cent}$ of the galaxies are classified as mergers/interactions, with a further $6\rm{\,per\, cent}$ possibly being in this 
category. The same percentages for the SFRNORM sample are $16\rm{\,per\, cent}$ and $6\rm{\,per\, cent}$, respectively. While these 
percentages are significant and hence identify the mechanism likely for triggering the starburst in somewhere between a fifth and a 
third of the cases, the fact nonetheless remains that the majority ($70-80\rm{\,per\, cent}$) of the galaxies must have had their 
starbursts triggered in some other way. Here our visual inspection of all our objects provided a possibly important additional clue. 
While many of our starburst galaxies were isolated galaxies (no near neighbour within the 1\,arcmin on a side SSS images used for our 
morphological classifications), the number that were in crowded fields with one or more very close neighbours was even more striking. 
Hence an important part of our analysis is to examine the `local' environments of our starburst galaxies and rigorously quantify whether 
such visual impressions are significant and determine whether the presence of a bright or faint neighbour is of any relevance in 
searching for other starburst trigger mechanisms (see Section 5). 
\begin{table*}
\caption{Morphological classifications for the SFRNORM and SFRTOT samples, based on visual inspection of SSS \bj\, and \rf\, images.}
\label{morph.table}
\begin{tabular*}{\textwidth}{@{\hspace{0.mm}}l@{\hspace{2.9mm}}c@{\hspace{2.9mm}}c@{\hspace{2.9mm}}c@{\hspace{2.9mm}}c@{\hspace{2.9mm}}c@{\hspace{2.9mm}}c@{\hspace{2.9mm}}c@{\hspace{2.9mm}}c@{\hspace{2.9mm}}c}
\hline
Sample & E/S0 & SE & SM & SL & Irr & Pec & Comp & Merger/ & Possible Merger/\\
&&&&&&&&Interact& Interact\\ \hline
\hline
SFRNORM&14 per cent&23 per cent&7 per cent&30 per cent&7 per cent&3 per cent&16 per cent&16 per cent&6 per cent\\ 
SFRTOT&29 per cent&35 per cent&11 per cent&23 per cent&0 per cent&2 per cent&0 per cent&28 per cent&6 per cent\\ \hline
\hline
\end{tabular*}
\end{table*}
\section{Luminosity Function}
\label{lumfun}
The luminosity function is an essential diagnostic tool for samples of galaxies. It encodes information about galaxy 
populations which may hold key clues to their evolution. For example, if it is found that our population
comprises only very luminous galaxies we could speculate that in order to trigger a starburst a major merger would
be required, whereas a sample of low luminosity galaxies are more likely to be affected by tidal interactions, hence may be triggered
by weaker interations.

We derive the \bj\, band luminosity functions, $\Phi(M)$, for our three (SFRNORM, SFRTOT and SFRTOT+) starburst samples and the \tdf\, 
sample using the Stepwise Maximum Likelihood (SWML) method described by \citet{efstathiou}. In order to minimize the effect of
varying completeness in the different \tdf\, fields, we set apparent magnitude limits of $14<$ \bj $< 19.2$
for all 3 samples. For the complete \tdf\, sample, the redshift was limited to $0.002< \rm z < 0.18$  to match the redshift limits of
our starburst samples. These limits restricted the initial samples to sizes of 311, 412, 239 and 108638 for the SFRNORM, 
SFRTOT, SFRTOT+ and \tdf\, catalogues, respectively. SSS \bj\, and \rf\, magnitudes were used to 
determine absolute magnitudes which were `K-corrected' and also `E-corrected' for the luminosity evolution of an 
average \tdf\, galaxy \citep{norberg}. 

\begin{figure*}
\center
\includegraphics[height=\linewidth,angle=-90]{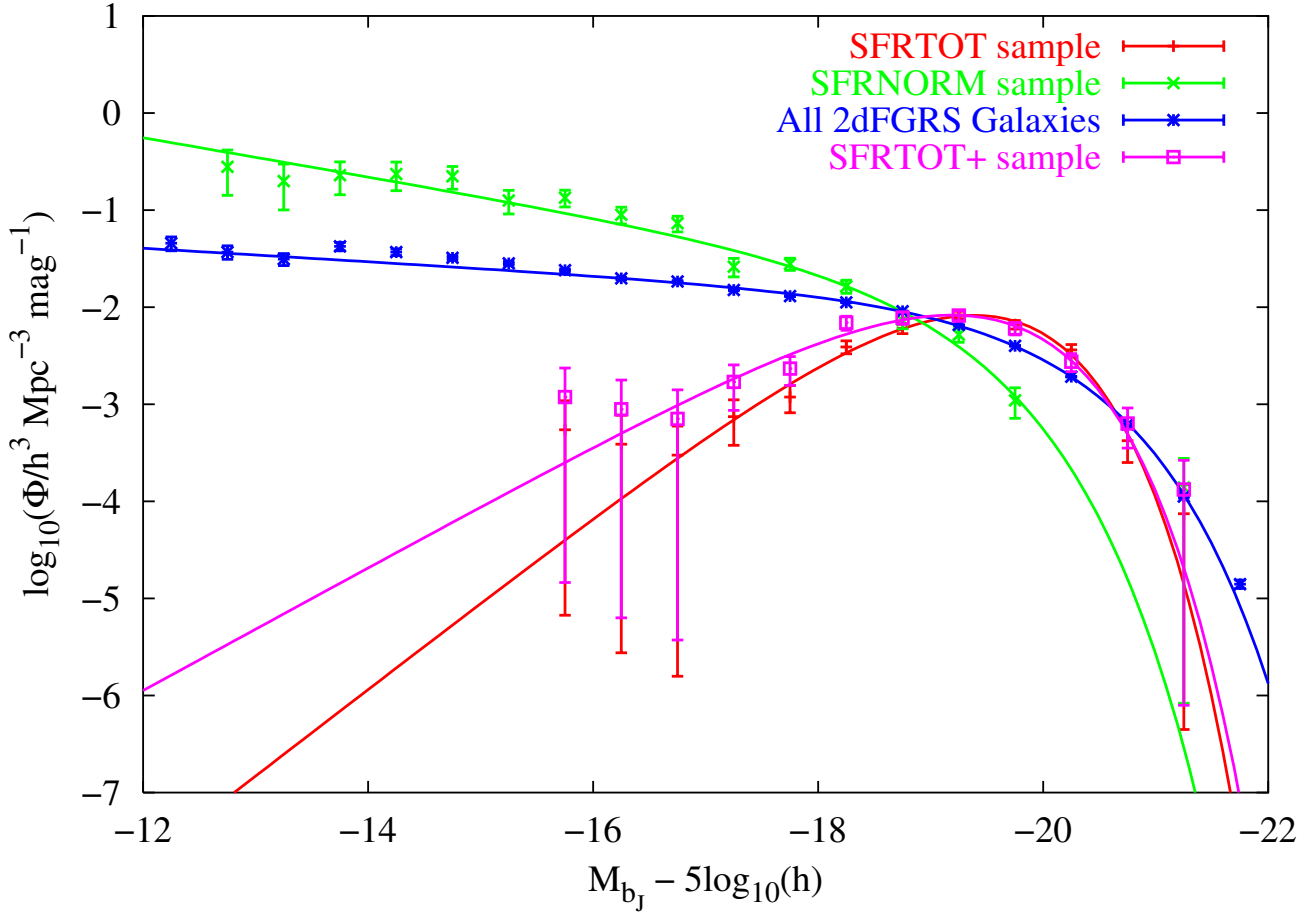}
\caption{The luminosity distributions and the fitted Schechter luminosity  functions for the SFRNORM ({\it green}), SFRTOT({\it orange}),
SFRTOT+({\it pink}) and \tdf({\it blue}) samples. The luminosity functions were derived using a stepwise maximum likelihood method and 
normalized to a constant source surface density of $\sigma = 124.6\, \rm{deg}^{-2}$. The errors are fractional Poissonian errors.}
\label{lumfunfig}
\end{figure*}

The SWML estimator requires normalization, hence we normalize each luminosity 
function to a constant source surface density, $\sigma$. To obtain $\sigma$, we used the parameters derived by 
\citet{norberg} in fitting a Schechter function to the \tdf\, catalogue in the absolute magnitude interval 
$-16.5 \le M_{{b}_{\rm{J}}} - 5 \rm{log}_{10}h \le -22$ (i.e., $M_{{b}_{\rm{J}}}^* - 5 \rm{log}_{10}h = -19.66$, $\alpha=-1.21$ and 
$\Phi^* = 1.61\times 10^{-2}\, h^3 \rm{Mpc}^{-3}$). We integrate the Schechter function using these parameters within the
redshift interval $0.002 \le z \le 0.18$ and magnitude interval $14 <$ \bj\, $<19.2$ to obtain 
$\sigma = 124.6\, \rm{deg}^{-2}$. This normalization allows us to investigate differences in shapes of the luminosity 
functions for each sample. The fractional errors presented for $\Phi(M)$ are Poisson $1/\sqrt{N}$ errors, where $N$ is the number of 
objects in the magnitude bin of interest. We used the least-squares method to fit Schechter functions to each of the \tdf, SFRNORM, SFRTOT
and SFRTOT+ luminosity functions, with the best fitting Schechter parameters presented in Table \ref{schechter.params}. Figure 
\ref{lumfunfig} shows the luminosity functions derived from the SFRNORM, SFRTOT,  SFRTOT+ and \tdf\, samples with best fitting Schechter 
functions overplotted. 

The appearance of the SFRNORM luminosity function is significantly different from that of the \tdf\, sample, 
and we quantify this using a chi-squared test. The chi-squared test is performed only where bins contain more than five objects 
(since the chi-squared test is not robust where bins contain less than 5 counts), and 
returns a vanishingly small probability that the SFRNORM luminosity function is consistent with that of the whole \tdf\, catalogue. This
also becomes clear on comparison of the SFRNORM and \tdf\, Schechter function parameters, with the SFRNORM sample showing both a fainter
$M^*_{{b}_{\rm{J}}} - 5\rm{log}_{10}(h)$ and steeper faint end slope ($\alpha$). It is clear that the SFRNORM sample is dominated by 
galaxies with $M_{{b}_{\rm{J}}} - 5 \rm{log}_{10}(h)>-18$ when compared to the overall \tdf\, sample. This portion of the luminosity 
function is generally occupied by dwarf and irregular type galaxies. The domination of low luminosity galaxies in the SFRNORM sample is 
not unexpected, since the normalized SFR is essentially a SFR per unit luminosity, hence the measurement is biased towards low 
luminosity galaxies. This bias does not detract from the physical importance of this sample; the SFRNORM sample just consists of a 
different class of starburst object, i.e., those that don't necessarily have prodigious SFR, but have SFR much higher then their 
quiescent SFR values.

The SFRTOT luminosity function also differs significantly from that of the \tdf\, luminosity function -- a chi-squared 
test conducted as outlined above shows a vanishingly small probability that the 2 luminosity functions are consistent, however, the 
SFRTOT sample differs in that it consists mainly of galaxies that are within $\pm1$\,mags of the value of  
$M_{{b}_{\rm{J}}}^*-5\rm{log}_{10}h$ derived by \citet{norberg} for the entire 2dFGRS sample. This can be understood by considering the 
morphological analysis in section \ref{morphologies}, where 69 per cent of the sample were classified as spiral type galaxies and 29 per 
cent as S0/E type. \citet{jerjen} presented type-specific luminosity funcitons where the S0 and spiral galaxy luminosity functions
are best described by a Gaussian function. Thus, as expected, our Schechter function fits have $\alpha>-1$, and the function behaves 
similarly to a modified Gaussian (see Table \ref{schechter.params}). The faint tail is supplied by a small number of low luminosity dwarf 
type galaxies where the luminosity function behaves like a Schechter function.

It should also be noted that a burst of star formation will increase the \bj\, magnitude of a galaxy, hence pushing it towards brighter 
magnitudes up the luminosity function. The galaxies at the faint end of the SFRNORM sample would most likely fall below the magnitude cut
off of the \tdf\, survey if they were not starbursting. 
\begin{table}
\caption{Schechter function parameters derived from a least-squares fit to the \tdf, SFRNORM, SFRTOT and SFRTOT+ samples. The errors are 
$\pm 1 \sigma$.}
\label{schechter.params}
\begin{tabular}{@{}l@{\hspace{1.5mm}}c@{\hspace{1.5mm}}c@{\hspace{1.5mm}}c@{\hspace{1.5mm}}c}
\hline
Sample & $M_{b_{\rm J}}^*-5\rm{log}_{10}(h)$ & $\alpha$ & $\Phi^*$ ($10^{-3}h^3 \rm{Mpc}^{-3}$) \\

\hline
\hline
\tdf\,&$ -19.65\pm0.02$&$-1.17\pm0.01$& $13.1\pm0.3$\\
SFRNORM &$-18.73\pm0.20$&$-1.50\pm0.07$&$27.2\pm7.6$\\
SFRTOT & $-18.51\pm0.07$&$1.24\pm0.14$& $13.8\pm1.4$\\
SFRTOT+&$-18.69\pm0.1$&$0.58\pm0.16$& $21.1\pm1.4$\\
\hline
\end{tabular}
\end{table}

\section{Local Environment}
\label{localenv}
We next examine the local environment of our starburst galaxies, and quantitatively assess whether it is of relevance to the starburst 
phenomenon. In doing so, we are particularly interested in the presence of near neighbour galaxies, the frequency of which was seen
to be quite high when morphologically classifying our starburst samples visually (see Section 3). We use the SSS to
extract catalogues of objects within 30\,arcmin radius from our starburst galaxies. The 
approximate magnitude limits for the photographic plates scanned to produce 
the SSS are \rf=21.5 and \bj=22.5. The objects in each catalogue are 
assigned a class parameter 1 through 4 based on morphological parameters and 
areal profile. In this classification scheme, galaxies are class=1, stars 
class=2, unclassified class=3 and noise class=4. During the analysis, we use 
only those catalogue members with SSS class=1, noting there may be some 
contamination by stars. 

We conduct experiments designed to test for correlations between the 
starburst galaxy and its environment on scales less than 1\,Mpc. During this analysis, we assume
all neighbours lie at the same redshift as the starburst and convert
angular separations to projected physical separations accordingly. 
Further to determining whether our starbursts have near neighbours, we would like to estimate the magnitude of the gravitational 
perturbation due to the presence of the nearest neighbour. Hence we define an arbitrary mass ratio whereby if the near neighbour is more
massive than one third of the mass of the starburst of interest, it constitutes evidence of some major interaction, whereas
if the mass of the nearest neighbour is less than one third the mass of the starburst, it constitutes a minor interaction. By assuming 
the mass-to-light ratio in the \rf\, band is the same for
the starburst and near neighbour, a mass ratio of 1:3 constitutes a magnitude difference of 1.2\,mag. We then define bright and faint 
near neighbours (ie. major and minor interaction), by comparing the neighbour 
\rf\, magnitude to that of the starburst magnitude corrected for the brightening due to the starburst, \rstarf\, 
(see Appendix A for more details on how \rstarf\, is derived), such that a faint neighbour has \rstarf+1.2$<$\rf$\leq 21.5$ 
and a bright near neighbour has \rf $\leq$ \rstarf+1.2. 
\begin{figure*}
  \begin{center}
    \begin{tabular}{llll}
      {\includegraphics[angle=-90,width=0.48\textwidth]{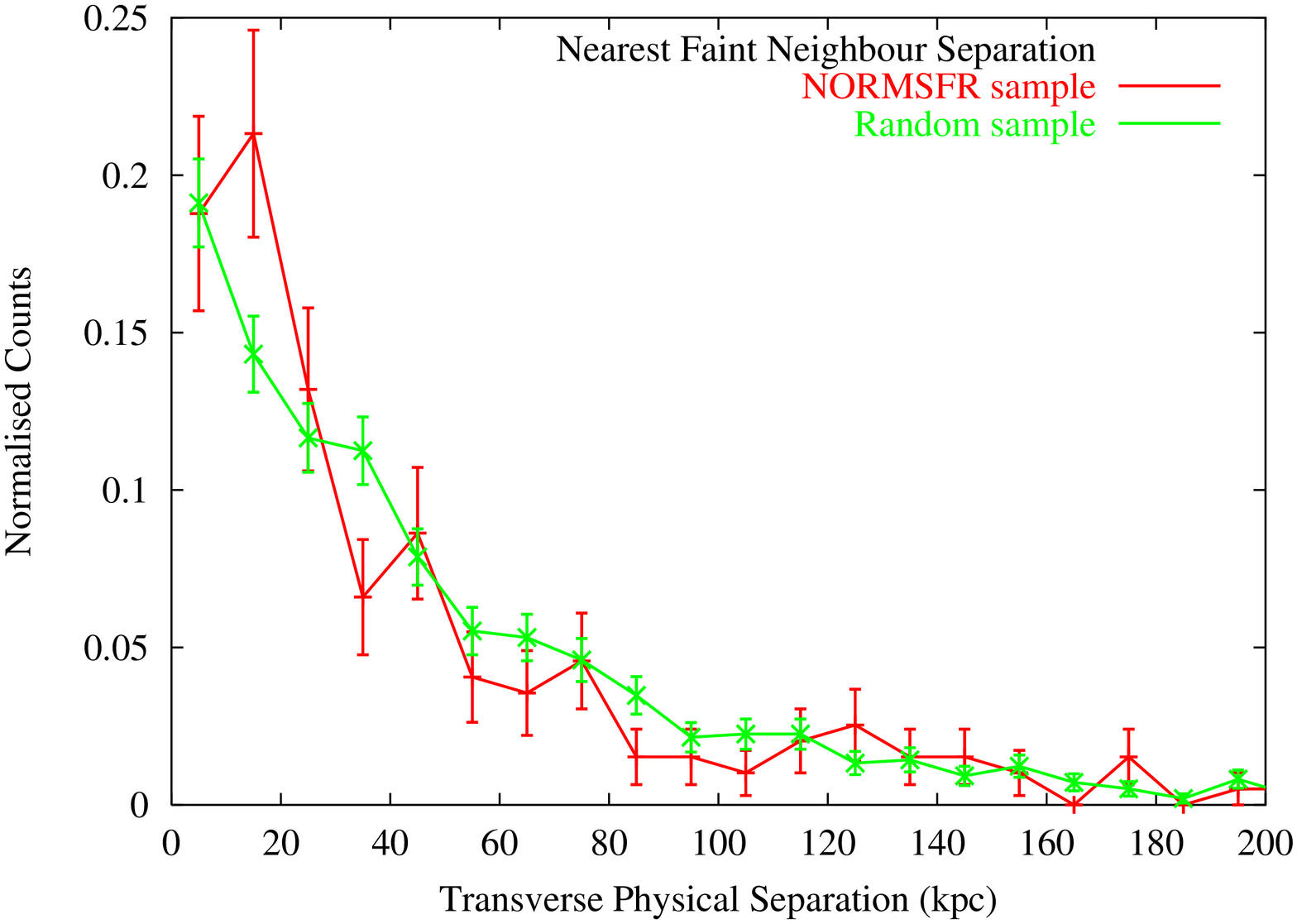}} &
      {\includegraphics[angle=-90,width=0.48\textwidth]{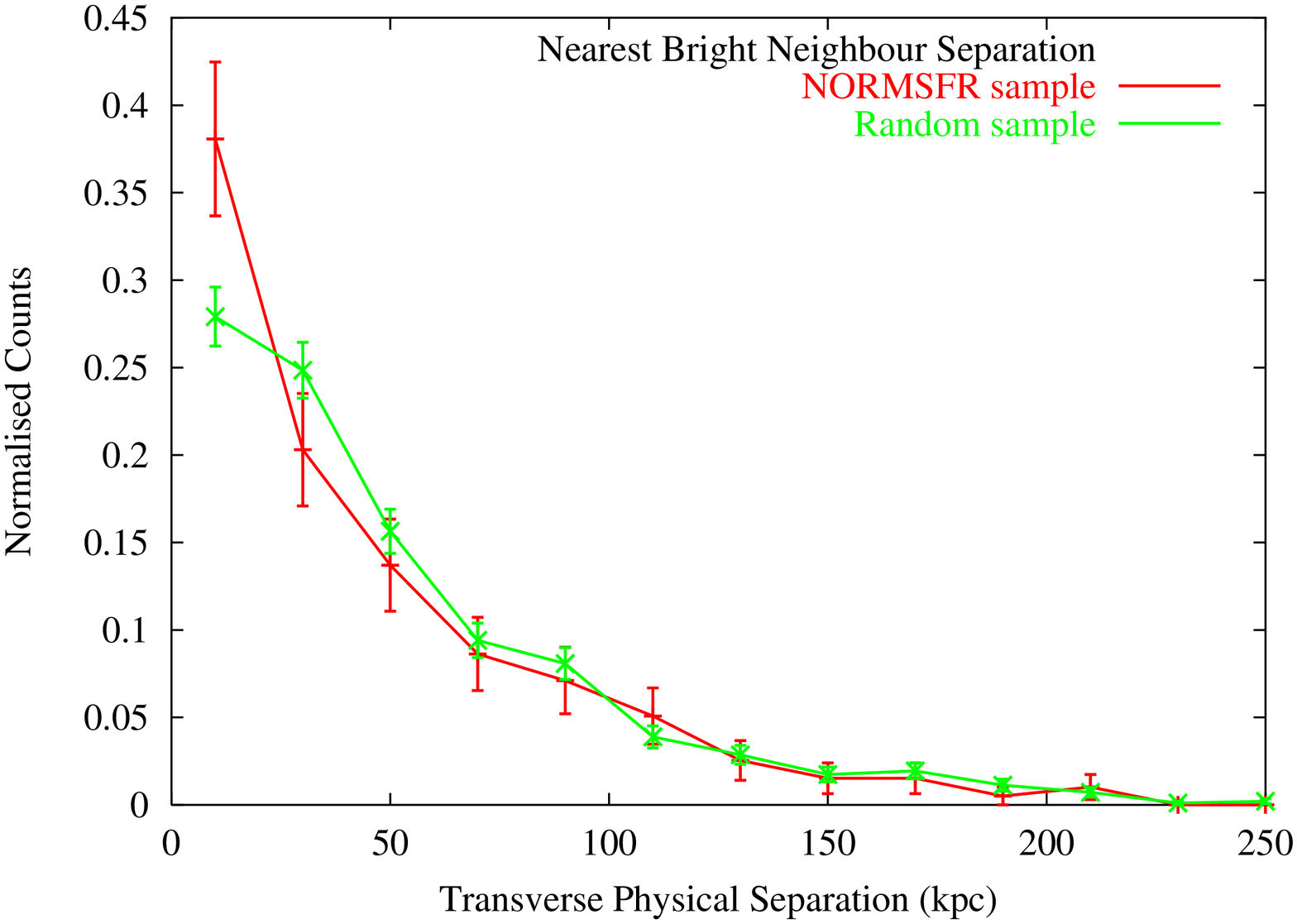}} \\
      {\includegraphics[angle=-90,width=0.48\textwidth]{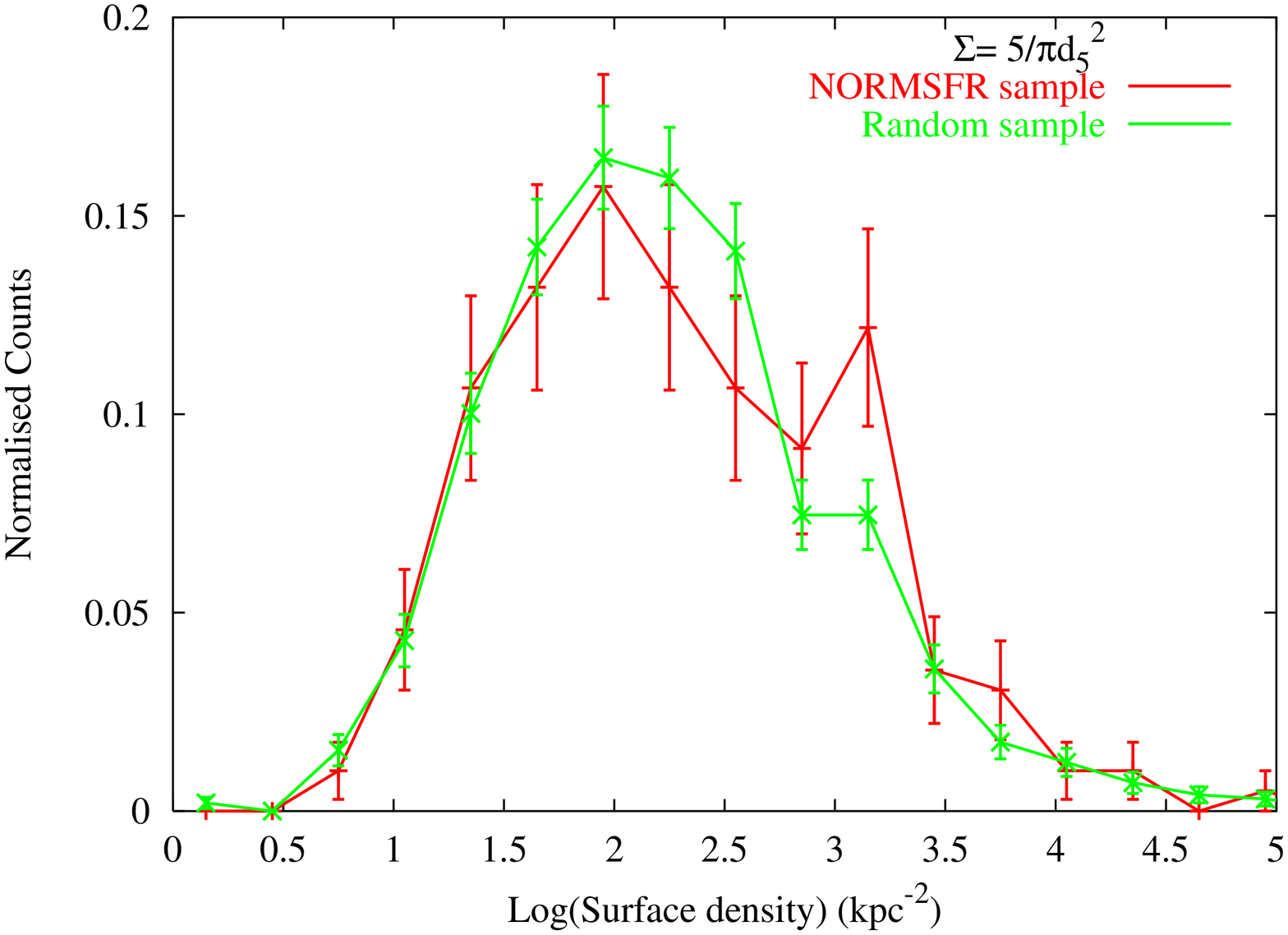}} &
      {\includegraphics[angle=-90,width=0.48\textwidth]{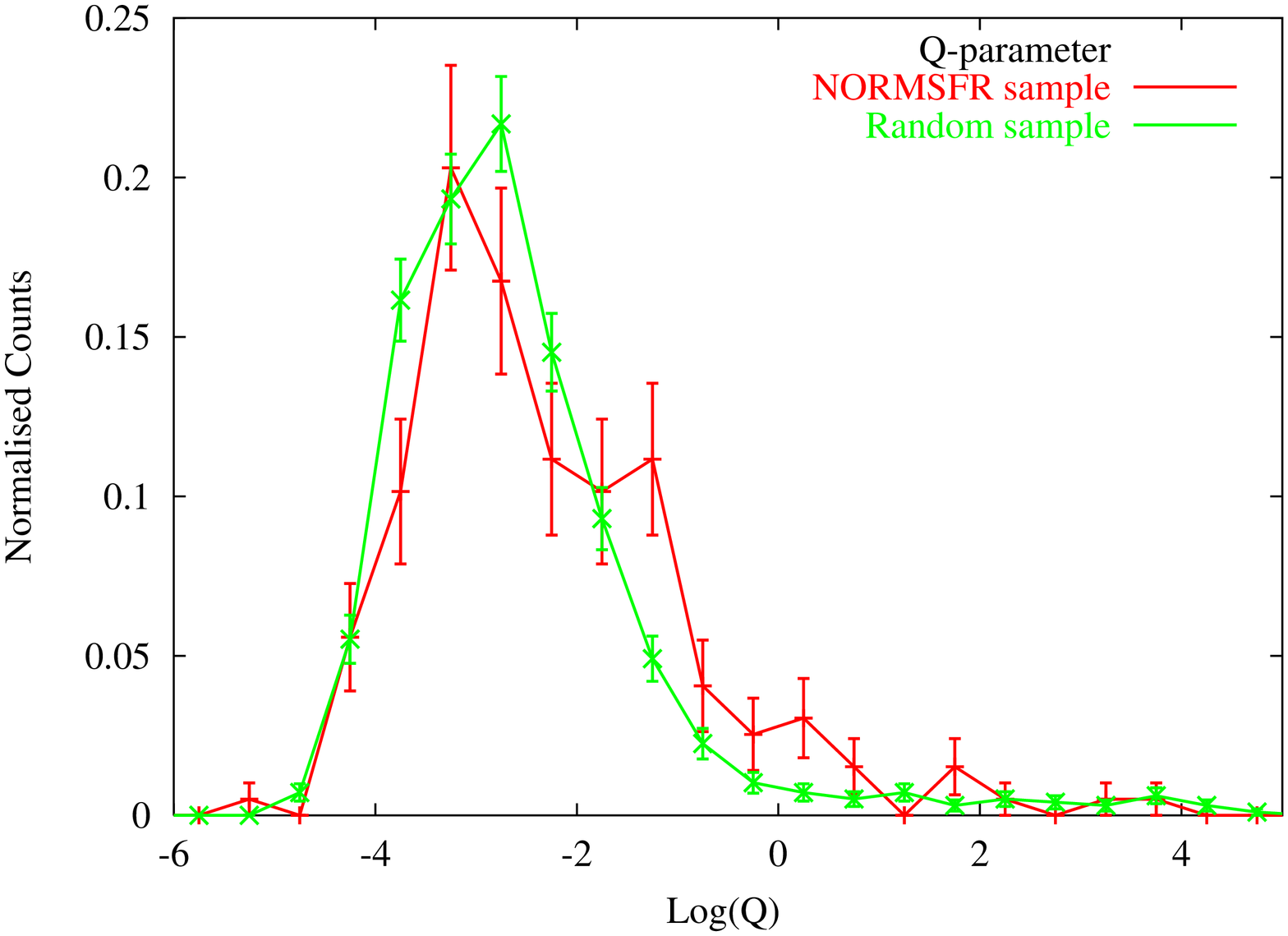}} \\
    \end{tabular}
    \caption{Near neighbour analysis results for the SFRNORM sample. The {\it upper left} plot shows the distribution of distances to 
the nearest faint neighbour, the {\it upper right} plot shows the distribution of distances to the nearest bright neighbour, the {\it 
lower left} plot shows the distribution of the surface density of the nearest five bright neighbours, and the {\it lower right} plot 
shows the distribution of the logarithm of the Q-parameters as described in Section \ref{localenv}. The KS probabilities (see text) are 
0.16, 0.015, 0.13 and 0.0017 for the faint near neighbour, bright near neighbour, surface density and Q-parameter, respectively.}
    \label{SFRNORM.nn}
  \end{center}
\end{figure*}
\begin{figure*}
  \begin{center}
    \begin{tabular}{llll}
      {\includegraphics[angle=-90,width=0.48\textwidth]{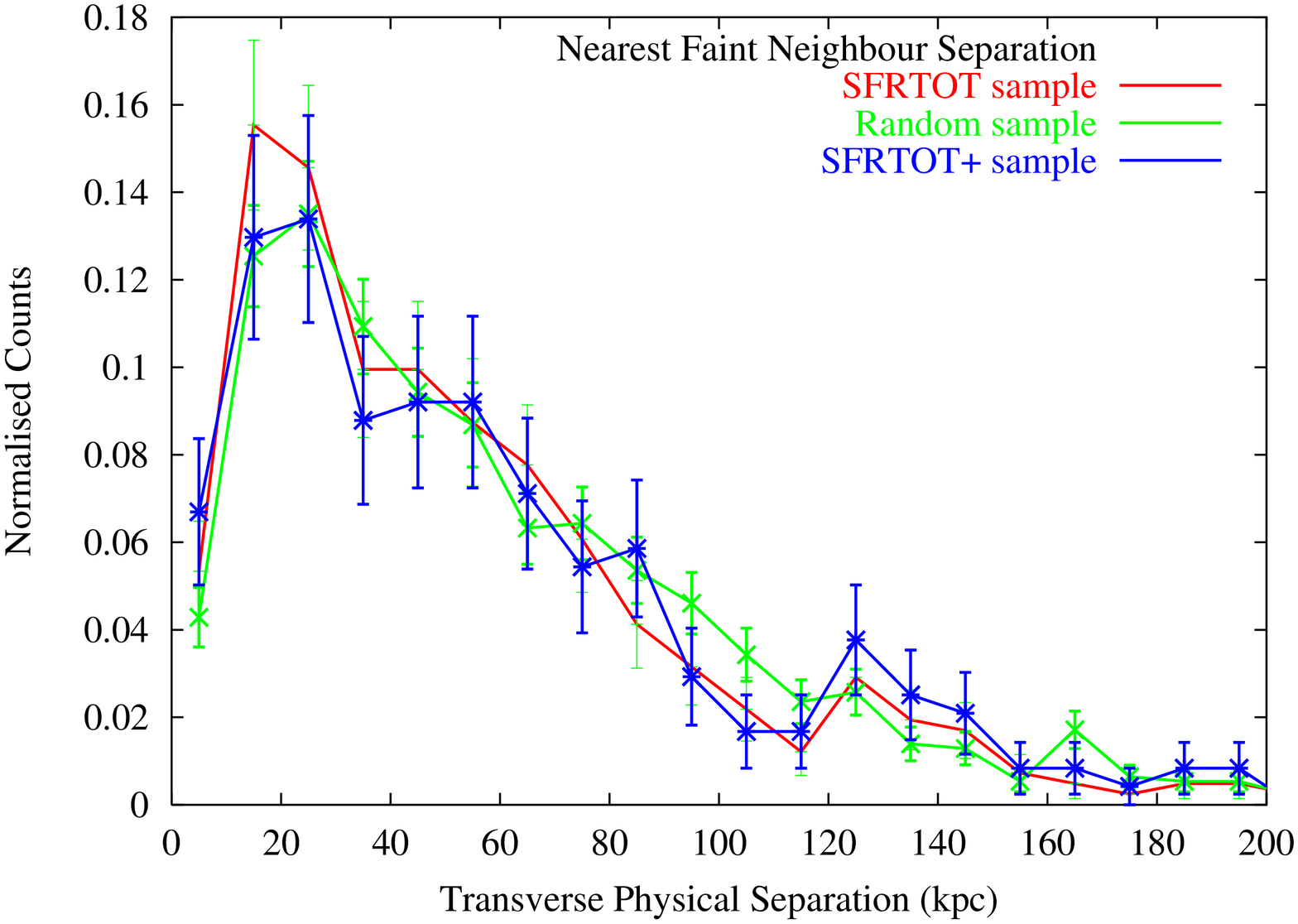}} &
      {\includegraphics[angle=-90,width=0.48\textwidth]{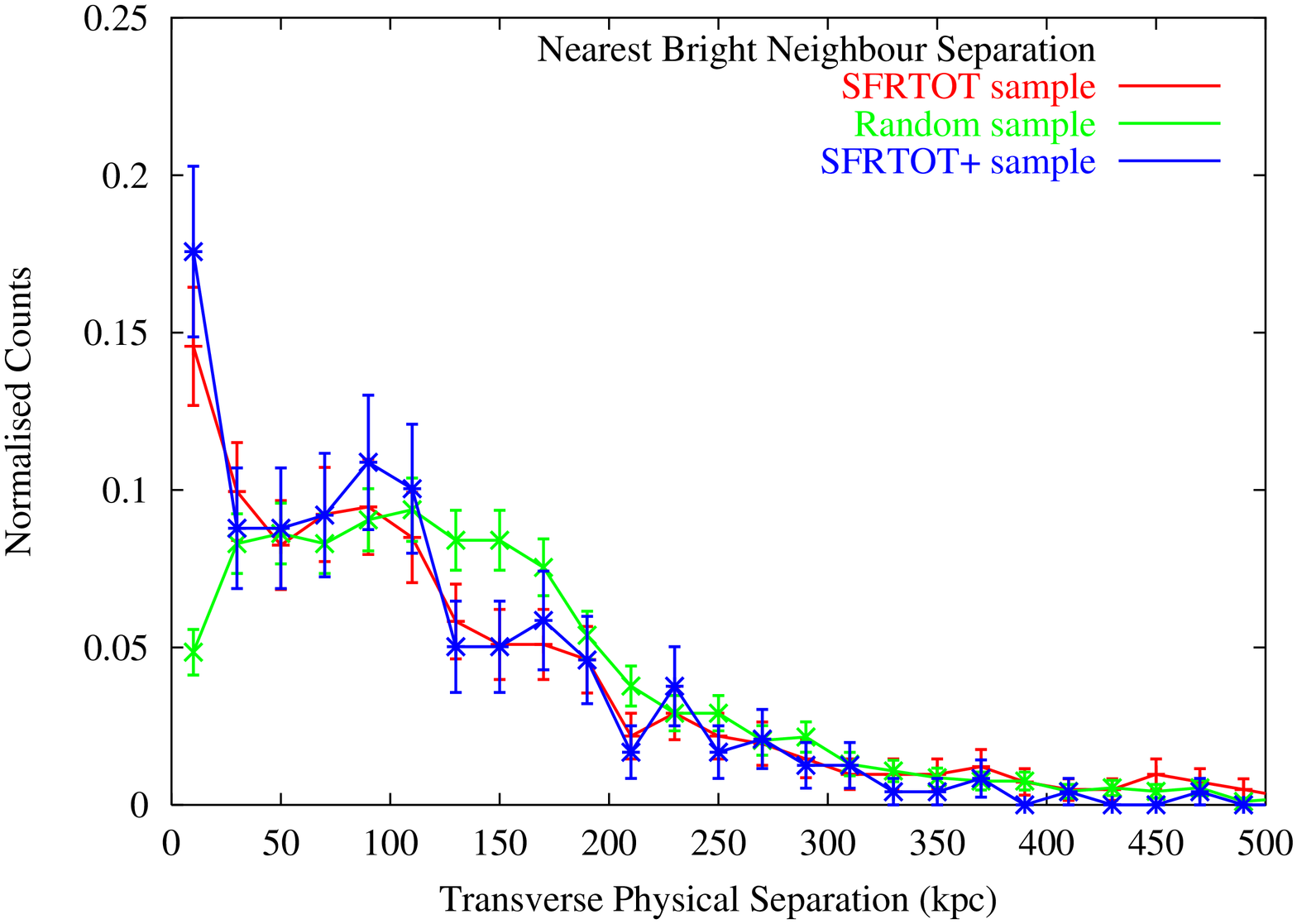}} \\
      {\includegraphics[angle=-90,width=0.48\textwidth]{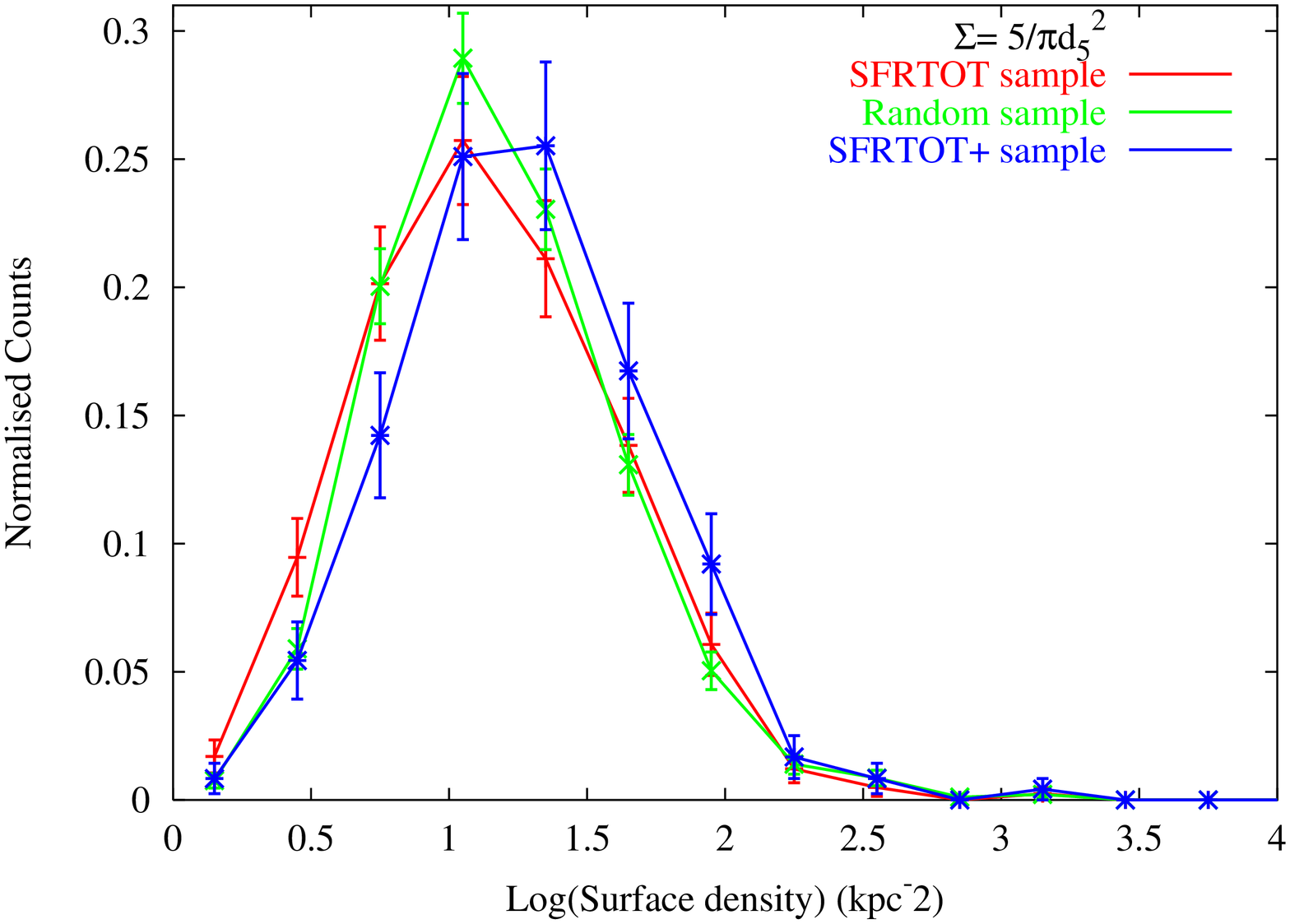}} &
      {\includegraphics[angle=-90,width=0.48\textwidth]{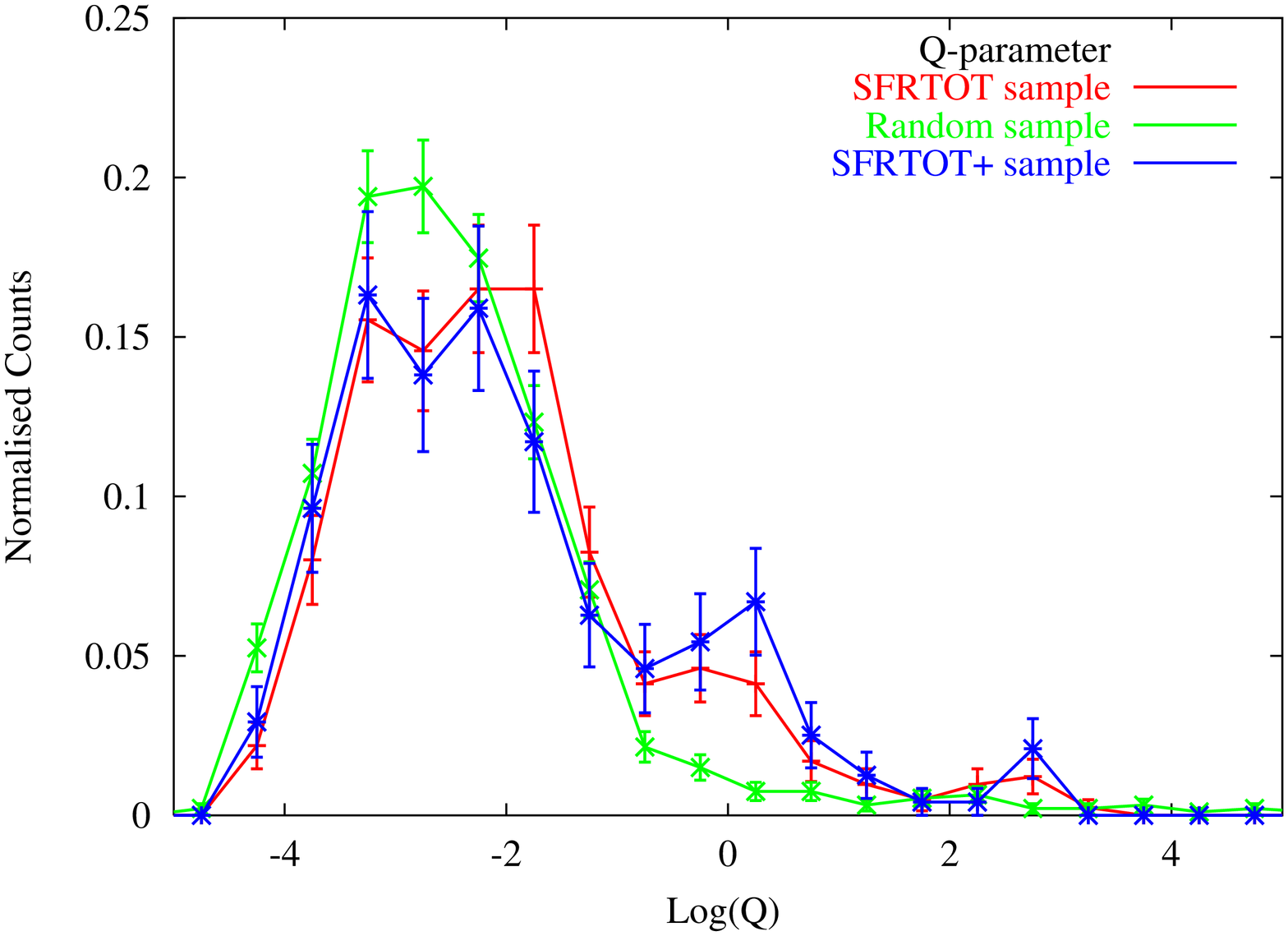}} \\
    \end{tabular}
    \caption{Same analyses as shown in Fig. \ref{SFRNORM.nn} for the SFRTOT ({\it orange}) and SFRTOT+ ({\it blue}) samples. The KS 
probabilities (see text) are 0.1 (0.76), $4.6\times 10^{-5}$ ($5.2\times 10^{-5}$), 0.3 (0.01) and $5.0\times 10^{-8}$ 
($6.7\times 10^{-5}$) for the faint near neighbour, bright near neighbour, surface density and Q-parameter, respectively where the SFRTOT+ 
values are given in brackets.}
    \label{SFRTOT.nn}
  \end{center}
\end{figure*}

As an alternative local environment indicator, we use the surface density of the five nearest bright 
neighbours, $\Sigma=5/\pi d^2_5$ where $d_5$ is the transverse physical distance to the fifth nearest bright neighbour. 

The final test is designed to estimate the interaction strength due to the 
nearest neighbour. We use the Q-parameter, which estimates the ratio of the 
tidal (${F}_{tidal}=GM_{NN}R_{SB}R^{-3}_{sep}$) to binding force 
(${F}_{bind}=GM_{SB}R^{-2}_{SB}$) \citep{dahari84} hence 
\begin{equation}
Q\propto\left(\frac{M_{NN}}{M_{SB}}\right)\left(\frac{D_{SB}}{R_{sep}}\right)^3
\approx\frac{{(D_{NN} D_{SB})}^{1.5}}{S^3}.
\end{equation}
Here, $M_{SB}/M_{NN}$ is the starburst/near neighbour mass and it is assumed that 
$M_{NN}\propto(D_{NN})^{1.5}$ and $M_{SB}\propto(D_{SB})^{1.5}$ \citep{rubin}, where $D_{NN}$($=2R_{NN}$) is the nearest neighbour 
semi-major axis physical diameter, whilst $D_{SB}$($=2R_{SB}$) is the semi-major axis physical diameter of the starburst of interest 
and $S \simeq R_{sep}$ where $S$ is the projected physical separation of the starburst and its nearest neighbour and $R_{sep}$ is the 
true three-dimensional separation. $D_{SB}$ and $D_{NN}$ are derived from the intensity weighted semi-major axis given in the SSS 
catalogues which are used in conjunction with the intensity weighted semi-minor axis to define an elliptical area for determination of 
isophotal magnitudes (see \citet{stobie} for more details). 

In order to gauge the significance of our results, for each test we compare 
our starburst sample results to that of a random sample containing 1000 galaxies. The selection of the 
random sample is crucial in order to negate systematic biases. Since we are
using projected information to seek correlations with physical associations, it 
is important the random sample has the same redshift distribution as the starburst
sample of interest. This ensures both samples have the same systematic contributions
due to foreground and background galaxies. The random sample should also have the same
magnitude distribution in each redshift bin. This is because we are using the \rf\,
magnitudes as a proxy for mass in an attempt to estimate the interaction strength.
Hence we select separate random samples for the SFRNORM and SFRTOT samples, each with 
redshift distributions determined by the starburst sample. The magnitude distributions
were chosen to correspond to the {\it brightening corrected} magnitude starburst 
distributions. The approximate \rf\, magnitude limit for the \tdf\, survey is \rf$\sim19$
(assuming on average \bj-\rf=0.5), hence the random sample contains only galaxies with
\rf\,$\leq 19$, whilst we only use starburst galaxies with brightening corrected magnitudes
\rstarf$\leq 19$. This additional constraint on the starburst magnitude produced samples of 197, 412 and 239 galaxies for the SFRNORM, 
SFRTOT and SFRTOT+ samples, respectively.

Figure \ref{SFRNORM.nn} shows the near neighbour results for the SFRNORM sample. The errors
presented are Poisson $\sqrt N$ errors. The Kolmogorov-Smirnov (KS) 
test was used to determine the probability that the random and starburst samples were drawn from the same 
parent distribution. The KS statistic measures the maximum value of the absolute 
difference in the cumulative probability distribution of two samples and determines 
the significance of the difference from the KS-statistic distribution. 
Hence, we determine a probability, p, that the KS statistic could exceed the observed value due to some random fluctuation in the data 
sets. A small value for p indicates two samples are not drawn from the same parent population. The p values determined for the faint 
nearest neighbour, bright nearest neighbour, surface density of bright neighbours and the Q-parameter are 0.16, 0.015, 0.13 and 0.0017 
respectively. These probabilities imply the bright near neighbour and Q-parameter distributions differ significantly (at the 98.5 and 
99.8 per cent levels, respectively) from the random distributions, with the bright near neighbour distribution having an excess of 
neighbours within $\sim 20$\,kpc, whilst the Q-parameter distribution has a higher proportion of galaxies with ${\rm log}(Q)>-1$. We find 
38 per cent of our SFRNORM galaxies have a bright neighbour within 20\,kpc, and upon cross correlating with the morphological analysis in 
Section \ref{morphologies} an additional 16 per cent of this sample are classified as merger/interacting or possible merger/interacting 
yet have no bright neighbour within 20\,kpc. 

Figure \ref{SFRTOT.nn} shows the near neighbour results for the SFRTOT and SFRTOT+ samples, with Poissonian 
errors presented. We used the KS test to derive p values for the faint nearest neighbour, bright nearest 
neighbour, surface density of bright neighbours and the Q-parameter of
.1(.76), $4.6\times10^{-5}$($5.2\times10^{-5}$), 0.3(0.01) and $5\times 10^{-8}$($6.7\times 10^{-8}$), 
respectively, where the results for the SFRTOT+ sample are shown in brackets. Again, we find the bright near neighbour and Q-parameter
distributions differ significantly from the random distributions (at $>99$ per cent level), where the bright near neighbour 
distribution shows an excess of near neighbours within $\sim 20$\,kpc and the Q-parameter distribution has
a higher proportion of galaxies with ${\rm log}(Q)>-1$. In addition to the 15 per cent for which we see a bright near neighbour within 
20\,kpc, we find 25 per cent are classified as merger/interacting or possible merger/interacting yet have no bright near neighbour within 
20\,kpc. Of the 22 per cent of the SFRTOT sample classified as E/S0, we find only one-quarter harbour a near neighbour within 50\,kpc, 
hence we have a significant population of isolated, starbursting E/S0 galaxies.

The fact that the $\Sigma_5$ distributions for the starbursts do not differ significantly from the overal random samples (similar to the 
findings of \citet{mateus04}), along with the statistically significant excess of bright near neighbours within 20\,kpc for both starburst 
samples, suggest that a near neighbour plays an important role in triggering a starburst. These near neighbour results are consistent 
with previous results showing a neighbour within 50\,kpc correlates with enhanced star formation rates \citep{nikolic04,lambas03,barton}. 
The results are also consistent with those of Freedman et al. (2006) who find that only galaxy pairs with small luminosity contrasts show
any trend of increased \ha\, emission at smaller projected radii. \citet{goto} also find a higher number of near neighbours within 50\,kpc 
for 266 E+A galaxies when compared to field galaxies, consistent with our results. The Q parameter confirms the importance of a near 
neighbour, with both starburst samples showing excess galaxies with ${\rm log}(Q)>-1$ compared to the random samples. Combining the near 
neighbour results with the morphological analysis in Section \ref{morphologies} {\it we conclude that galaxy-galaxy interactions are 
important in triggering 40--50 per cent of the starbursts in our samples}. Further to this, we note that there is no excess of faint near 
neighbours in either sample, implying that the interaction required to trigger a massive starburst needs to be significant.

\section{Large scale environment}
\label{largescale}
Here we investigate the environments and clustering properties of our starburst galaxy samples on more global scales, that is those 
larger than $1$\,Mpc.
\subsection{Membership of groups from the 2PIGG catalogue}
The 2PIGG galaxy groups catalogue \citep{ekea} was compiled from the \tdf\, catalogue using a friend-of-friends percolation algorithm. 
The catalogue was drawn from the contiguous SGP and NGP regions. For the purposes of our analysis, we restrict our attention to groups at
redshifts $z<0.12$, since above this limit interloper contamination increases, whilst the fraction of the observed group luminosity 
drops below half the total group luminosity. There are two catalogues each for the NGP and 
SGP, the first contains the \tdf\, galaxy information along with a flag of 0 for ungrouped or 
the group ID number if the galaxy is linked to a group, the second contains information 
pertaining to group properties such as velocity dispersion, the unweighted number of group 
members and the rms projected galaxy separation. The catalogues are publicly available and 
can be obtained at http://www.mso.anu.edu.au/2dFGRS/Public/2PIGG/. 

We would like to ascertain whether our starburst galaxies reside in particular types of groups when compared to a sample of 5000 random 
galaxies selected from the parent \tdf\, catalogue such that their redshift distribution is the same as that of the starburst sample. There
are a number of methods to determine group properties from the 2PIGG catalogues. These include dynamical mass 
estimates, the number of galaxies within each group and total group luminosity. \citet{ekeb} 
showed that the weighted total group luminosity is the most robust estimator of group size, since the dynamical estimators have large 
errors for the small groups, whilst the number of observed members residing in a higher redshift group can be significantly lower than 
the actual number due to the survey magnitude limit. Here we have used the weighted observed group \bj\, band luminosity 
defined by \citet{ekeb} as 
\begin{equation}
L_{obs,{b}_{\rm{J}}}=\sum_{i}^{n_{gal}} w_i L_{i,{b}_{\rm{J}}},
\end{equation}
where $w_i$ is the weight correcting for the incompleteness of the \tdf, $L_{i,{b}_{\rm{J}}}$ 
is the \bj\, luminosity of the ith group member and $n_{gal}$ is the number of galaxies assigned to 
the group. From the observed weighted group luminosity, the total weighted group luminosity was 
determined by dividing by the incomplete Gamma function 
$\Gamma(\alpha+2,L_{min}/L^*)/(\Gamma(\alpha+2))$. This allows extrapolation from the lowest 
luminosity galaxy detectable at the group redshift, $L_{min}$, 
to zero luminosity, to correct for group members lying below the survey magnitude limit. The correction is derived from the integral 
of the Schechter luminosity function in the luminosity range 0 to $L_{min}$, using global Schechter parameters given by \citet{ekeb}
as ($M_{{b}_{\rm{J}}}^*$,$\alpha$)=(-19.73, -1.18). The luminosity function is known to vary amongst groups of different masses, however
using luminosity function parameters derived on a group mass basis changes the total group luminosities by no more than $\sim 10$ 
per cent \citep{ekeb}.  

\begin{figure*}
  \begin{center}
    \begin{tabular}{cc}
      {\includegraphics[angle=-90,width=0.48\textwidth]{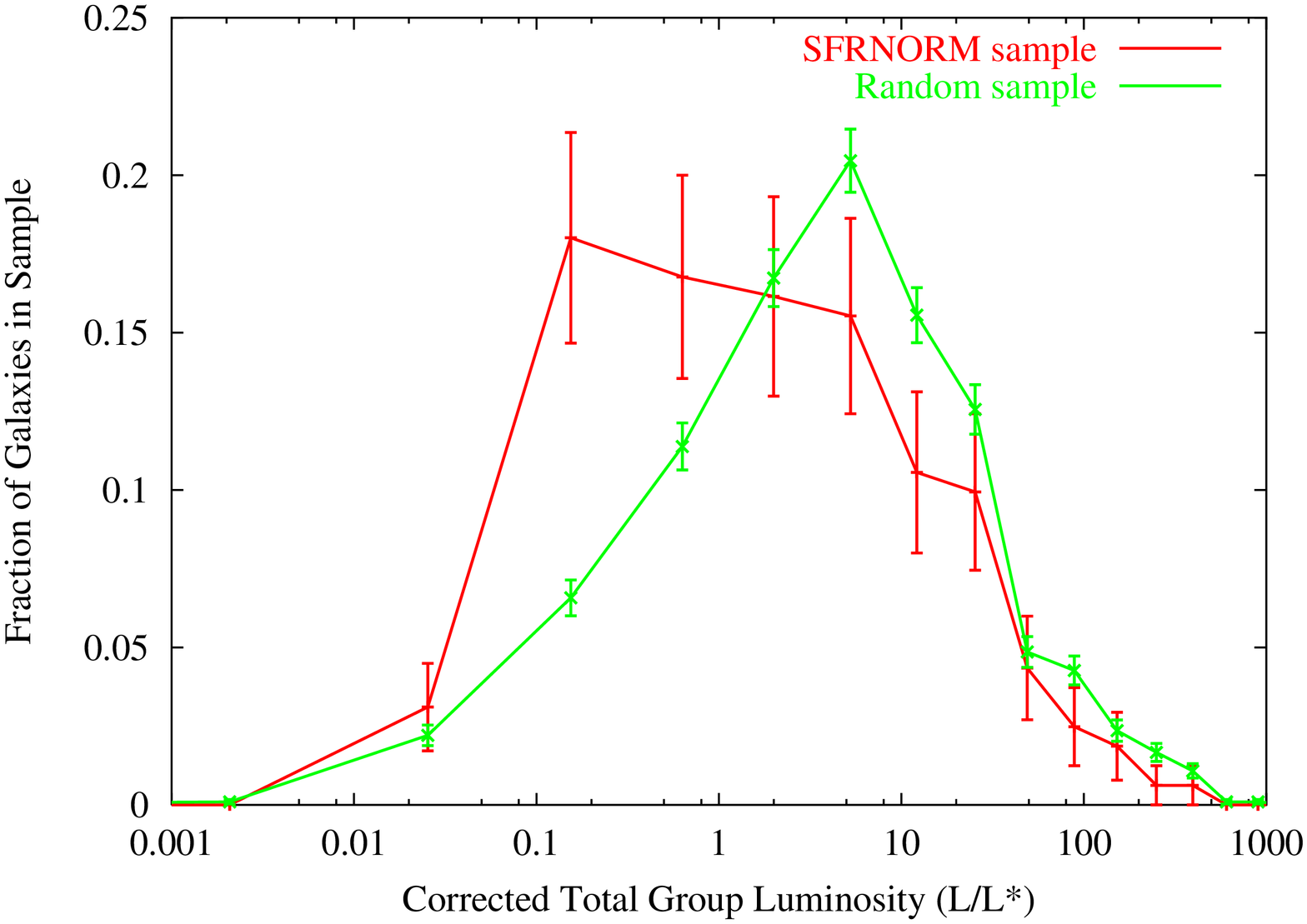}} &
      {\includegraphics[angle=-90,width=0.48\textwidth]{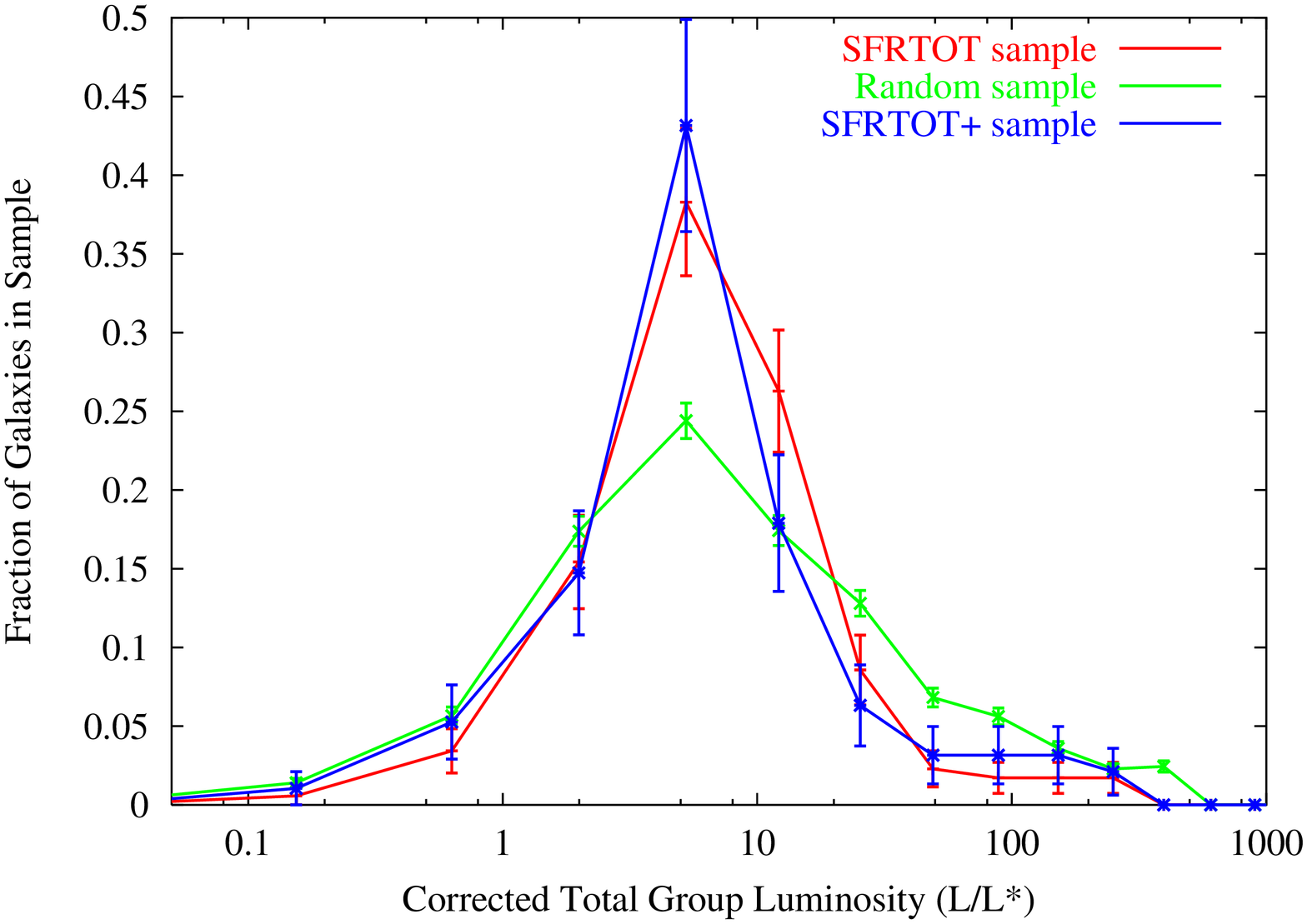}} \\
    \end{tabular}
    \caption{The distribution of corrected total group luminosities for
the SFRNORM sample ({\it left}) and SFRTOT sample ({\it right}) compared to the same distributions for the associated random samples (see
text for details). Error bars show Poissonian $\sqrt N$ values, where $N$ is the number of galaxies in each bin. Both starburst samples 
differ from their respective random samples at greater than 99 per cent significance.}
    \label{grp.plots}
  \end{center}
\end{figure*}
In Figure \ref{grp.plots} we plot the distributions of the corrected total group luminosities for the
starburst and random samples. We find 52, 58 and 56 per cent of the galaxies in the SFRNORM, SFRTOT, and SFRTOT+ samples reside 
in groups, respectively. Notably, these percentages are no different to those for the 2dFGRS galaxy population as a whole, with both 
random samples having 56 per cent of their galaxies residing in groups. Both the chi-squared and KS tests were used to determine the 
probability that the starburst and random samples were drawn from the same parent distribution with both tests giving probabilities of 
less than 1 per cent of this occurring for each starburst sample. The distribution of luminosities for the SFRNORM sample are skewed to 
fainter and presumably poorer groups compared to the random sample, whilst the SFRTOT sample is strongly peaked at around 8$L^*$, ie. 
medium sized groups. 

\subsection{Membership of rich clusters}

\citet{propris} have compiled a list of galaxy clusters from the Abell, APM and EDCC catalogues
lying within the \tdf\, regions. Redshifts, velocity dispersions and cluster centroids were
measured for the majority of the clusters using the \tdf\, data. We use both redshift information
and cluster-centric distance to determine cluster membership for our starburst galaxies, using only those 
clusters with measured redshifts. The velocity dispersion was
used to determine the radius, $r_{200}$, at which the density of the cluster exceeds that of the critical density 
by a factor of 200 on the assumption the cluster is an isothermal sphere (see Carlberg et al. 1997). $r_{200}$ is an
approximation of the virial radius of the cluster and is defined by:
\begin{equation}
r_{200}=\frac{\sqrt3 \sigma_v}{10H(z)},
\end{equation}
where $\sigma_v$ is the velocity dispersion, $z$ is the cluster redshift and 
$H(z)=H_0\sqrt{\Omega_m(1+z)^3+\Omega_{\Lambda}}$. For those clusters without a measured $\sigma_v$
we used the average $\sigma_v$ for the cluster sample in determining $r_{200}$. We determine the peculiar velocity of the
starburst with respect to the cluster, 
\begin{equation}
v_{pec}=\frac{c((1+z_{pec})^2-1)}{((1+z_{pec})^2+1)},
\end{equation}
where $z_{pec}= (z_{sb}-z)/(1+z)$ is the 
non-cosmological redshift contribution of the starburst galaxy to its observed redshift, $z_{sb}$, due to its peculiar 
velocity within the cluster. We determined $r$, the projected physical distance of the starburst from the cluster centroid as 
$r=\Delta \theta \chi(z)/(1+z)$ where $\chi(z)$ is the radial comoving distance to the cluster and $\Delta \theta$ is the angular 
separation of the cluster centroid and the starburst galaxy.
Those starburst galaxies with $r \le r_{200}$ and $v_{pec} \le 3\sigma_v$ are classed as cluster
members, whilst those with $r_{200} < r \le 2r_{200}$ and $v_{pec} \le 3\sigma_v$ are classed as lying in the cluster outskirts and 
those with $2r_{200} < r \le 4r_{200}$ and $v_{pec} \le 3\sigma_v$ are classed as being within the infall regions of the cluster.

Using the above definitions, we find for the SFRNORM sample that 3, 10 and 19 per cent of starbursts lie within the 
cluster, outskirt and infall regions, respectively, compared to 8, 8 and 21 per cent respectively, 
for our random sample of 5,000 galaxies. The same analysis performed on the SFRTOT (SFRTOT+)  sample produces 5 (5), 
6 (7) and 20 (22) per cent within the cluster, outskirt and infall regions, respectively, compared to 9, 10 
and 24 per cent for our random sample of 5,000 galaxies. Hence, we 
conclude that the starburst galaxies within our sample are not preferentially found within clusters, with $\sim 30\rm{\,per\, cent}$ 
total found within $4 \times r_{200}$ Mpc of a cluster centroid. 

\subsection{Residence within large-scale overdensities}
\label{overdense}

As another method of quantifying the large scale structure surrounding our starburst galaxies, we measured the overdensities, $\delta$, 
within comoving spheres of radii $1\leq r_0\leq 15$\,Mpc in 1\,Mpc intervals. Specifically, we used the comoving distance of each 
surrounding galaxy to transform its right ascension, declination and redshift into 3-D cartesian coordinates, making no attempt to 
correct for peculiar velocity distortions in redshift space, and measure its physical separation from the starburst galaxy of 
interest. For each comoving sphere centred on a starburst galaxy, we add up the number of galaxies 
in the \tdf\, catalogue within $r_0$ giving the number of observed galaxies within the sphere, $N_{obs}$. The number of galaxies 
expected in each comoving sphere in the absence of clustering, $N_{exp}$, was determined using random mock catalogues generated by the 
publically available \tdf\, mask software written by Peder Norberg and Shaun Cole (see http://www2.aao.gov.au/2dFGRS/). Again, using 3-D
cartesian coordinates, we measure the physical separation between the mock catalogue galaxy and the starburst of interest. We obtain 
$N_{exp}$ by adding up the number of mock galaxies surrounding each starburst galaxy within $r_0$. The average 
overdensity within each comoving sphere is calculated as follows:
\begin{equation}
\overline{\delta(r_0)}=\frac{1}{n_{sb}} (\sum_{i=1}^{n_{sb}} 30\frac{N_{obs}}{N_{exp}}-1)\label{dens},
\end{equation}
where $n_{sb}$ is the number of starburst galaxies used in the analysis. The random mock catalogue was generated such that the sample 
size is 30 times that of the \tdf\, catalogue out to z=0.18 with a limiting magnitude of \bj=19.45, hence the multiplicative factor of 
30 in eqn. \ref{dens}. The larger sample size for the mock random catalogue was used in order to minimize the noise in the denominator 
caused by low number statistics.
\begin{figure*}
  \begin{center}
    \begin{tabular}{ll}
      {\includegraphics[width=0.48\textwidth]{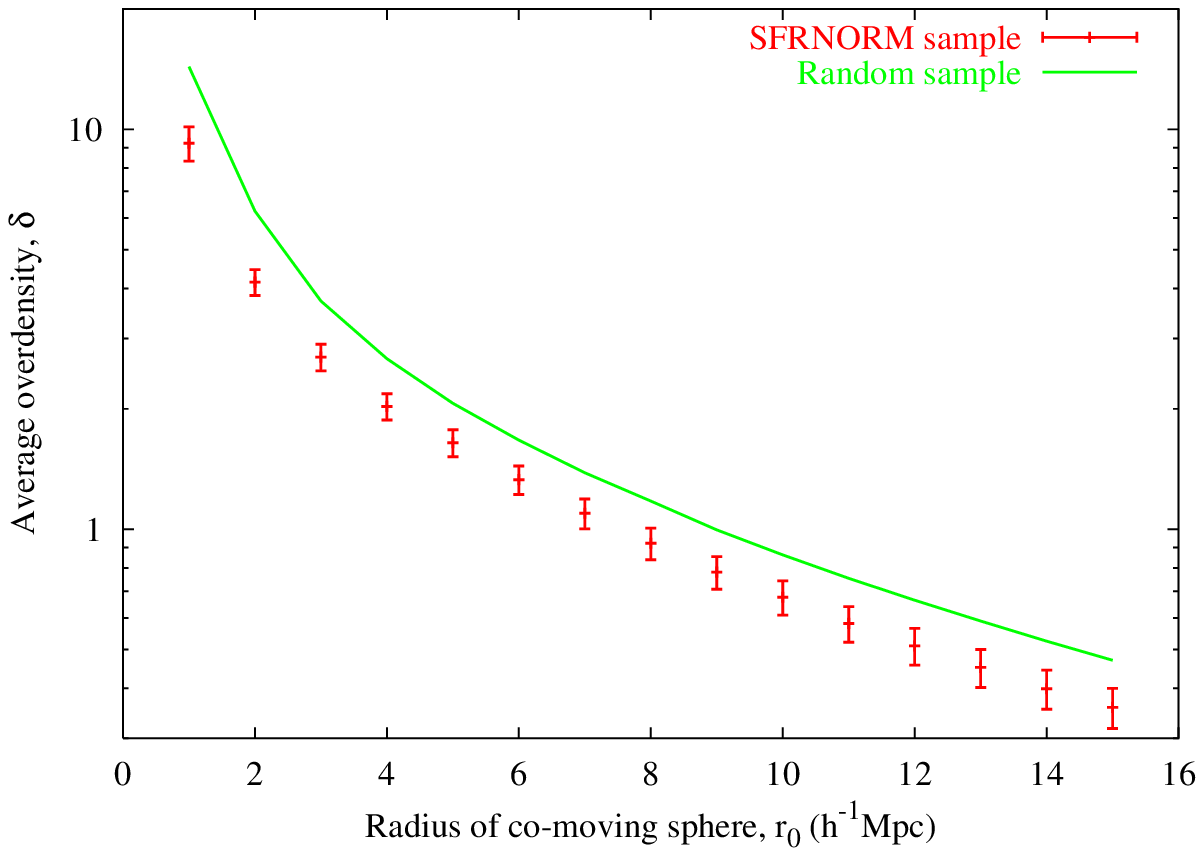}} &
      {\includegraphics[width=0.48\textwidth]{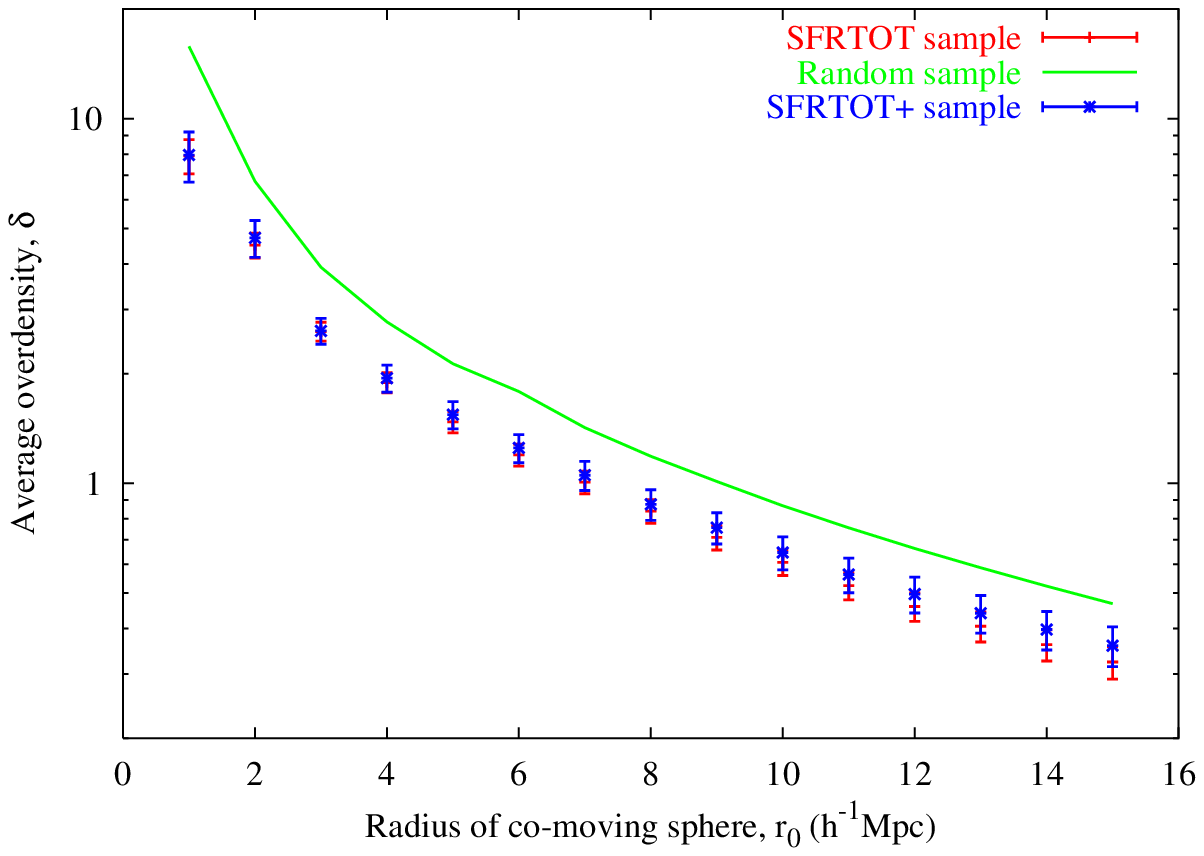}} \\
    \end{tabular}
    \caption{The average overdensity surrounding each starburst galaxy within comoving spheres of radius $r_0$ obtained by
comparing the number of galaxies in the \tdf\, within $r_0$ with that of a mock random catalogue. The analysis was repeated on a 
sample of 10,000 random \tdf\, galaxies selected such that the redshift distribution is the same as the starburst sample of interest.
The figure on the {\it left} shows the results for the SFRNORM sample, whilst that on the {\it right} shows the results for the SFRTOT 
and SFRTOT+ samples. The error bars represent the variance in the overdensity distribution; for the random samples these are negligible 
and hence not presented.}
    \label{overdens.plots}
  \end{center}
\end{figure*}

We compared to a random sample selected from the \tdf\, to have the same redshift distribution as the starburst sample of interest, 
such that redshift dependent biases are minimized. We used only the SGP and NGP regions for this analysis, the 
results of which are presented in Figure \ref{overdens.plots}. The error bars presented are the variance, $\sigma/\sqrt{n_{sb}}$, of the 
overdensity distribution where $\sigma$ is the standard deviation of the distribution of mean overdensities and $n_{sb}$ is the number 
of starburst galaxies in the sample. The results clearly indicate that both 
starburst samples reside in lower density environments than the randomly selected \tdf\, galaxies on all scales from 1\,Mpc to 15\,Mpc. A
common measure of environment is the overdensity within spheres of radius $8h^{-1}$\,Mpc. On this scale, we measure average overdensities
of 0.92 ($\pm0.08$) and 0.84($\pm0.06$) for the SFRNORM and SFRTOT samples, respectively, which are both consistent (within errors) with 
that measured by \citet{blake} for the \tdf\, E+A sample. For the random samples we measure 1.17($\pm0.02$) and 1.19($\pm0.02$) for the 
SFRNORM and SFRTOT random samples, respectively. Despite using a different method to measure the number of galaxies expected in each 
sphere in an unclustered Universe, the results are, within the errors, consistent with the value measured for a random sample by 
\citet{blake}.

\subsection{Spatial cross-correlation function}

The spatial cross-correlation function measures the clustering properties of galaxies by estimating whether a particular type of galaxy
is biased to inhabit high mass or low mass haloes when compared to the bias of another type of galaxy. If we approximate a linear bias
for the galaxies in the starburst samples, $b_{sb}$, then their spatial auto-correlation function, $\xi_{sb,sb}$, as a function of
spatial separation, $r$, takes the form
\begin{equation}
\xi_{sb,sb}(r)=b_{sb}^2\xi_{m,m}(r),
\end{equation}
where $\xi_{m,m}(r)$ is the spatial auto-correlation function of the underlying mass density field. 

Due to the relatively low number of galaxies in our starburst samples, shot noise dominates the auto-correlation function,
hence we measure the cross-correlation function, $\xi_{sb,g}$, of the starburst galaxies with the rest of the \tdf\, catalogue. In the
linear bias approximation,
\begin{equation}
\xi_{sb,g}(r)=b_{sb}b_g\xi_{m,m}(r),
\end{equation}
where $b_g$ is the bias factor for an average \tdf\, galaxy. We can estimate the relative bias of the starbursts to the \tdf\, galaxies 
by measuring the auto-correlation function for the \tdf\, galaxies, $\xi_{g,g}(r)=b_{g}^2\xi_{m,m}(r)$, such that
\begin{equation}
\frac{b_{sb}}{b_g}=\frac{\xi_{sb,g}(r)}{\xi_{g,g}(r)}.
\end{equation}

We have measured all correlation functions in redshift space, making no attempt to correct for peculiar velocities. The 
cross-correlation function is measured by comparing the cross-pair counts of the starburst samples with the full \tdf\, catalogue
and an unclustered random distribution with the same redshift distribution and selection mask as the \tdf\, 
sample, whilst in order to minimize fluctuations due the selection function of the starburst galaxies, we measure the cross pair counts of
the \tdf\, with an unclustered random sample generated with the same redshift distribution as the starburst sample of interest. The random
distributions were generated using the same publicly available \tdf\, mask software as in Section \ref{overdense} and contain the same 
number of galaxies as the real data samples they are simulating. We use a modified Landy-Szalay estimator 
\citep{blake06} for the cross-correlation function, $\xi_{sb,g}(s)$, which is given by
\begin{equation}
\xi_{sb,g}(s)=\frac{N_{sb,g}(s)}{N_{Rsb,Rg}(s)}-\frac{N_{sb,Rg}(s)}{N_{Rsb,Rg}(s)}-\frac{N_{g,Rsb}(s)}{N_{Rsb,Rg}(s)}+1, 
\label{crosscorr.eqn}
\end{equation}
where $N_{sb,g}(s)$ is the starburst-\tdf\, pair cross count, $N_{sb,Rg}(s)$ is the starburst-simulated unclustered random \tdf\, 
sample pair cross count, $N_{g,Rsb}(s)$ is the \tdf\,-simulated unclustered random starburst sample pair cross count,
$N_{Rsb,Rg}$ is the cross pair count for the two random samples and the $s$ 
denotes a redshift space separation. Each of the $N_{sb,Rg}(s)$, $N_{g,Rsb}(s)$ and $N_{Rsb,Rg}$ are measured by generating 10 random
catalogues and taking the average of the 10 measurements. In order to estimate the auto-correlation function for the \tdf\, galaxies, 
$\xi_{g,g}$, for comparison, we use the Landy \& Szalay estimator \citep{Landy93}, given by

\begin{equation}
\xi_{g,g}(s)=\frac{N_{g,g}(s)}{N_{Rg,Rg}(s)}-\frac{N_{g,Rg}(s)}{N_{Rg,Rg}(s)}+1,
\label{autocorr.eqn}
\end{equation}

where $N_{g,g}$ is the auto-pair count for the \tdf\, galaxies, $N_{g,Rg}$ is the \tdf\,-random pair cross count and $N_{Rg,Rg}(s)$ is 
the random-random pair auto count. 

\begin{figure}
\center
\includegraphics[angle=-90,width=\linewidth]{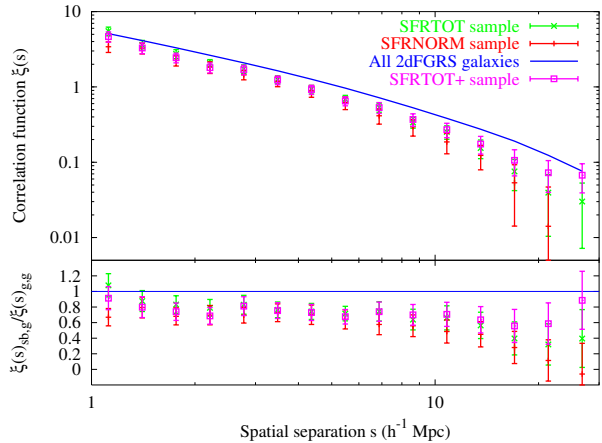}
\caption{Spatial correlation function measurements for the SFRNORM, SFRTOT, SFRTOT+ and \tdf\, samples. The plotted errors are derived 
using a jack-knife re-sampling technique, as described in the text. The lower plot shows the ratio of $\xi_{sb,g}(s)/ \xi_{g,g}(s)$ for 
the SFRNORM, SFRTOT and SFRTOT+ samples.}
\label{crosscorr}
\end{figure}

We include the contiguous NGP and SGP strips in our analysis, but not the random fields since the variance in the correlation function 
is large due to edge effects. The results are displayed in Figure \ref{crosscorr}. We do not assign Poisson error bars
here, since they are known to underestimate the true variance of the estimators in equations \ref{crosscorr.eqn} and \ref{autocorr.eqn}
by a significant amount \citep{Landy93}. The errors presented are estimated using the `jack-knife' approach whereby the NGP and SGP 
strips are divided into forty regions and the correlation function estimation is repeated forty times, each time leaving out one region. 
The error for each separation bin is then estimated by multiplying the resulting standard deviation across the forty subsamples by 
$\sqrt{40}$.

The results are presented in Figure \ref{crosscorr} where $\xi_{g,g}(s)$ and $\xi_{sb,g}(s)$ are plotted for the SFRNORM, SFRTOT, SFRTOT+ 
and the \tdf\, catalogues. We also plot the ratio of $\xi_{sb,g}(s)/ \xi_{g,g}(s)$.  The SFRNORM, SFRTOT and SFRTOT+ samples are much less
clustered than the entire \tdf\, sample, on all scales (apart from the smallest separation bin for the SFRTOT and SFRTOT+ samples, which 
are consistent with the \tdf\, sample within errors) implying that $b_{sb}<b_g$. In fact, the average ratio of $b_{sb}/b_g$ is 0.53, 0.69 
and 0.73 for the SFRNORM, SFRTOT and SFRTOT+ samples, respectively. These results are consistent with the \citet{blake} `average Balmer' 
E+A catalogue, which is found to be somewhat less clustered than the \tdf\, catalogue, although it is noted their sample size is only 50, 
hence the results are tentative. The results are also consistent with those of \citet{madgwick03}, who find that active galaxies, defined 
by the $\eta$ parameter \citep{madgwick02}, are less clustered than passive galaxies out to real space separations of $10 h^{-1}$Mpc.

\section{Summary and discussion}
\label{discussion}
We have selected two samples of starburst galaxies, each containing 418 galaxies, using different methods for deriving the star 
formation rate from the \ha\, emission line. The selection methods are complementary, in that selecting starburst galaxies based on 
$W_{H\alpha}$ alone is known to be biased towards low mass galaxies (our SFRNORM sample), whereas the selection of galaxies by measuring
the total absolute star formation rate selects a sample of more luminous, and presumably more massive galaxies (our SFRTOT sample). 
Hence, we cover a broad spectrum in mass with our two starburst samples. We study the environments and morphologies of the galaxies in 
these samples in an effort to ascertain mechanisms which may be important in triggering a burst of star formation, and in truncating 
the star formation leading to the E+A type spectrum.
\subsection{Summary of results}
The morphological analysis of our starburst galaxies indicates:

(i) Both starburst samples contain galaxies covering the full range of Hubble types, with early types (E/S0, SE) being more dominant in 
the SFRTOT sample than the SFRNORM sample.

(ii) 20--30 per cent of the starburst galaxies, irrespective of how they are selected, exhibit signs of major merger/interaction or tidal 
activity. 

(iii) 29 per cent of the SFRTOT sample are classified as E/S0 and of these early types 69 per cent show no evidence for merging/interaction
nor do they harbour a bright near neighbour within 20\,kpc.

We have conducted the same environmental analyses on a sample of random \tdf\, galaxies, selected such that systematic effects relevant
to the particular analysis are minimized. Comparison of our starburst samples with these random samples shows:

(i) The luminosity function of the SFRNORM sample has a significantly steeper faint-end ($M_b>-18$) slope compared to the \tdf\, 
catalogue. As noted above, selection based on $W_{H\alpha}$ alone is biased towards selecting low mass galaxies, hence we should expect 
a high number of low luminosity galaxies in our sample. In contrast, the SFRTOT sample shows a dearth of galaxies at $M_b>-18$, and is 
dominated by $L^*$ and brighter galaxies. The dearth of fainter galaxies is presumably caused by the fact that the low luminosity galaxies 
don't have high {\it absolute} SFRs, but do have high SFRs compared to their quiescent states. Indeed, the majority of galaxies in the 
SFRNORM sample have SFRTOT values $<10$\,\msolar\,$yr^{-1}$.

(ii) Both samples have a significant overdensity of bright near neighbours within 20\,kpc. The SFRTOT 
sample also shows an excess number of galaxies experiencing tidal forces due to a near neighbour. This excess is also 
present in the SFRNORM sample, although at a lower level of significance. For the local galaxy surface density and projected distance to 
faint near neighbour studies, we find the distributions for the starburst samples are consistent with that of a random sample. 

(iii) On larger scales, we find those SFRNORM galaxies residing in groups preferentially inhabit low luminosity groups, whilst 
a higher fraction of those SFRTOT galaxies residing in groups inhabit average luminosity groups. We find our 
starbursts do not preferentially inhabit clusters of galaxies. Upon measuring the overdensity within comoving spheres on scales 
1--15\,${h}^{-1}$\,Mpc we find that starbursts lie in underdense regions compared to random samples on all scales. The cross-correlation 
function also shows that our starburst galaxies are less clustered on scales larger than 2\,${h}^{-1}$\,Mpc compared to the \tdf\, 
galaxies.

(iv) We find throughout the analysis that the SFRTOT+ sample shows no significant differences from that of the SFRTOT sample.

From this analysis, we can conclude that starburst galaxies are less clustered and hence less biased on large scales when compared to 
passive galaxies and appear to be mainly found within the field rather than clusters of galaxies. Local environment appears to be the
key determinant in triggering a starburst, in the form of mergers and tidal interactions. The analysis of the types of groups our 
starbursts inhabit is consistent with the merger/interaction hypothesis when we consider that the SFRNORM sample is made up of primarily
low luminosity/mass galaxies with presumably low velocity dispersions, whilst the SFRTOT sample consists of more luminous/massive 
galaxies which should have higher velocity dispersions. Mergers are rare in environments where the 
galaxy velocity dispersion is much higher than the stellar velocity dispersion, hence if mergers are the
dominant mechanism triggering a starburst then we would expect to find low mass starbursts in low mass groups, and 
high mass starbursts in intermediate mass groups, consistent with the results presented here.

It is also clear from our analysis that a large fraction of our starbursts show no evidence for merger/interaction, nor do they have a 
bright close near neighbour. Hence, some form of galaxy interaction can not have triggered the starburst. What mechanism is triggering 
the starburst in these galaxies? Also of note is that 22 per cent of the SFRTOT sample is classified as E/S0 with no signs of merger 
activity, whilst our near neighbour analysis shows only 25 per cent of these undisturbed E/S0 galaxies harbour a bright near neighbour 
within 50\,kpc. It may be that there is some merger activity, but either our criterion is too strict to define the galaxy as a 
merging/interacting system, or the SSS imaging is simply too poor to resolve any evidence of this. It may also be that there has been 
some tidal interaction with a near neighbour some time ago, and the near neighbour is now at some distance from the starburst, hence does
not show up in our near neighbour analyses. For example, if a starburst was triggered by a galaxy interaction 500\,Myrs ago and the near
neighbour galaxy was traveling at 100\,km\,${\rm s}^{-1}$, then the near neighbour would be some 50\,kpc distant from the starbursting galaxy. 
This oversimplified example serves to show how evidence for a merger or tidal interaction may be missed in our analyses. There is also
the possibility that these galaxies are in the later stages of swallowing a gas-rich faint dwarf galaxy which is no longer distinguishable
from the host galaxy. However, if these starbursts were not triggered by some interaction, then an internal mechanism may be required to 
explain the burst. Detailed follow-up spatially resolved spectroscopy and high resolution imaging is critical in order to determine the 
nature of the triggering mechanism in these galaxies.  

\subsection{Comparison with \tdf\, E+A galaxies}
Concerning starburst galaxies as E+A progenitors, it is important to ask if our samples are likely to evolve into E+A galaxies
based on the evidence presented here and in \citet{blake}. It is important to note at the outset here, that \citet{blake} found only 56 
`true' E+A galaxies, whilst each of our starburst samples contain 418 galaxies despite probing a smaller volume. Perhaps only a small 
fraction of our starburst galaxies will go through an E+A phase.
\subsubsection{Luminosity functions}
Comparison of the E+A and SFRNORM luminosity functions shows a
marked difference at the faint end ($M_{b\rm J}-5 {\rm log}_{10}h < -18$) from which it might be inferred that our SFRNORM sample 
galaxies will not evolve into E+A galaxies. However, if our sample is predominantly dwarf starbursts which undergo periodic bursts of 
star formation with a duty cycle of ~10\,Myr, then we would expect post-starburst dwarf galaxies would display the characteristic E+A 
spectrum on very short time scales. 
Coupled with the fact that a large fraction of our SFRNORM sample are close to the magnitude limit of the \tdf\, survey and 
would not be detected in their un-brightened, quiescent state, we expect low luminosity E+A galaxies would be very rare in the \tdf\, 
survey. Indeed, this is seen in the \citet{blake} survey. Thus, it is unlikely that our SFRNORM sample will evolve into the E+As
observed by \citet{blake}.

Comparison our SFRTOT luminosity function with that of \citet{blake}
`average Balmer' sample (EW $\sim$ 4.5\,\AA), shows that the two samples differ at 97.7 per cent significance level in the range 
$-17.75>M_{b_{\rm J}}-5 {\rm log}_{10} h>-20.75$ where the difference occurs due to a higher number of 
$M_{b_{\rm J}}-5 {\rm log}_{10} h=-19 \pm 1$ galaxies in the 
SFRTOT sample. Taking this at face value, it would seem that the our starburst galaxies can not be progenitors to E+A galaxies.
However, since we are measuring the \bj\, band luminosity function, we would expect the luminosity function of
a starburst sample to be significantly different from that of a sample of quiescent galaxies with the same mass distribution. This
is because the radiation emitted by the starburst is dominated by a population of young, massive O-type stars which emit predominantly at
blue wavelengths, and hence the \bj\, luminosity of a starburst galaxy is significantly enhanced compared to its quiescent state. After 
the starburst is truncated and the galaxy enters the E+A phase, its \bj\, luminosity gradually fades back towards its quiescent value. 
This evolution in \bj\, magnitude means the luminosity function is not a particularly good discriminator for the two samples.
The k-band luminosity function would provide a much better discriminator here, since it is less affected by the light emitted from 
starburst regions. Therefore, the difference in the \bj\, luminosity functions alone is not enough to rule out an evolutionary link 
between our starbursts and the E+A galaxies of \citet{blake}.
\subsubsection{Local environment}
\citet{blake} find no statistically significant differences between their E+A samples and random \tdf\, samples in their near 
neighbours analysis, whereas here we find clear differences, particularly for the bright nearest neighbours within $\sim 20$\,kpc. This
is also somewhat expected, since simulations have shown that the E+A phase generally occurs at a late stage in mergers when the cores of 
the two galaxies can no longer be distinguished \citep{bekki01b}, although \citet{goto} find an enhancement in the number of near 
neighbours in their E+A sample. Also, \citet{blake} find a significant portion of their E+A sample
exhibit tidal tails and merger morphologies and the sample is made up of predominantly E/S0 galaxies, hence they conclude that major 
mergers are an important formation process for E+As. We then conclude that a sub-sample of our starbursts are likely to evolve into E+A 
galaxies after the rapid star formation has consumed the available gas.
\subsection{Caveats}
It is worth noting some caveats to our conclusions. The first is the \tdf\, fibre size, which is 2\,arcsec in diameter. This corresponds
to 4\,kpc at z=0.11, hence the covering fraction of the fibre is in general much smaller than the average  galaxy size. Studies of 
aperture effects have shown that measured properties such as corrected SFR and metallicity are affected when the covering fraction of a 
2\,arcsec fibre is $<20\rm{\,per\, cent}$, which corresponds to a redshift of z=0.06 for galaxies with similar properties to those in the 
Near Field Galaxy Survey \citep{kewley05}. Thus we should expect our SFRTOT sample to have SFR overestimated for $z<0.06$. This effect is
somewhat nullified by the fact that we select our sample from the most extreme galaxies per redshift bin. The effect should also be 
negligible in the SFRNORM sample due to two main reasons. The first being this sample is dominated by dwarfs, and indeed the distribution
of diameters derived from SSS parameters peaks at $\sim 3$\,kpc, hence the covering fraction drops below $20\rm{\,per\, cent}$ at around 
z=0.035, affecting some 35 per cent of our SFRNORM sample. The second is our SFRNORM measurements are based on $W_{H\alpha}$, which is a 
relative quantity for which we see minimal trends out to $z=0.18$. The second caveat comes from the fibre 
positioning in the \tdf\, survey. The fibres are most likely to be placed at the approximate centre of the galaxy, thus if the covering 
fraction is small,  we only receive 
light from the central regions meaning we are biased to selecting nuclear starburst galaxies. This may mean we miss particular types of 
starburst galaxies which have non-nucleated starbursts, eg., those with bursts triggered by ram-pressure from the ICM where the burst is 
expected in the galaxy outskirts, or those where the burst is triggered by gas sloshing. Follow up long-slit spectroscopy of a sub-sample
of starburst galaxies will allow us to decipher how much these aperture affects should bias our results.

\subsection{Outlook}
Clearly, from this study it can be concluded that a significant fraction of starbursts are triggered by some form of merger/interaction 
event. However, equally important are those galaxies that show no evidence for interactions, and hence what might have triggered their 
starburst. Follow up studies of these galaxies including high resolution optical imaging (for morphological analysis), spatially resolved 
spectroscopy (for mapping out the star forming regions across the galaxy and measuring stellar kinematics) and HI imaging (for constraining
gas dynamics) will allow us to determine whether the triggering mechanism is internal or external in these starbursts. 

\section{{Acknowledgments}}
We acknowledge financial support from the Australian Research Council through a Discovery Project grant held throughout the course if 
this work. MSO was also supported by an Australian Postgraduate Award. This paper would not be possible without the \tdf\, data collected
at the Anglo-Australian Observatory (AAO). We thank \tdf\, team and the staff at the AAO for making this study possible.

\appendix
\section{Correcting for the Brightening in the \rf\, Magnitude due to a starburst}
During the near neighbour analysis, we assumed a constant mass to light ratio for
the starburst and near neighbour galaxies. This assumption is subject to errors
caused by the brightening of the \rf\, magnitude due to the burst of star formation
occurring in the starburst, hence, we attempt to correct for this using 
the stellar population synthesis models of 
Bruzual \& Charlot (2003). Briefly, we superpose a 300\,Myr burst of constant 
star formation onto an underlying 11\,Gyr old stellar population which has an 
exponentially decaying star formation rate with an e-folding time of 
$\tau=3$\,Gyrs beginning 1\,Gyr after the Big Bang. There are two
main inputs for the models which we can derive from our spectra -- the
burst strength (ie. the ratio of the mass of stars formed in the burst to 
the underlying stellar mass) and the metallicity. 

The burst strength was determined from the $W_{H\alpha}$) of the starburst sample of interest (for 
the SFRNORM the median $W_{H\alpha} = 145$\,\AA\, and for the SFRTOT sample the median 
$W_{H\alpha} = 67$\,\AA. The $W_{H\alpha}$ correlates strongly with the `b' 
parameter \citep{scalo} which measures the ratio of the current star formation 
rate to the mean past star formation rate \citep{ken94}. Using Table 1. from \citet{ken94}, we estimate the equivalent b 
parameters for the median $W_{H\alpha}$ are 2.5 and 1 for the SFRNORM and SFRTOT samples, respectively. 
Multiplying the `b' parameter by the ratio of the assumed ages of the starburst (300\,Myr) and 
underlying stellar population (11\,Gyr), we can derive the burst strength. We derive burst 
strengths of 7 and 3 per cent for the SFRNORM and SFRTOT samples, respectively.

To determine the metallicity, we obtain the gas phase oxygen abundances using the EWs of OIII(4959,5007), 
[OII][3737) and \hb\, to derive the $\rm{R}_{23}$ parameter (see Kobulnicky et 
al. 2003) which in turn is used to determine the oxygen abundance, 12+log(O/H) 
(see Kobulnicky et al. 1999). The degeneracy between the $\rm{R}_{23}$ 
parameter and its relation to the oxygen abundance is broken by defining metal
rich galaxies as those with log([NII]$\lambda6584$/\ha)$\geq-1$ and metal poor
as those with  log([NII]$\lambda6584$/\ha)$<-1$ (see Lamaraille et al. 2004 
and references therein for a more thorough discussion). Eqns. (8) and (9) of 
\citet{kob99} are then used to determine the 12+log(O/H). We find mean oxygen 
abundances of 20 and 60 per cent of the solar value for the SFRNORM and SFRTOT 
samples, respectively. Hence we use the corresponding metallicities in the BC03
models. We find that, on average, a galaxy in the SFRNORM sample is brightened 
by $\sim0.83$ mag and a galaxy in the SFRTOT sample is brightened by 
$\sim0.44$ mag in the \rf\, band due to a starburst. Thus, for the SFRNORM sample \rstarf=\rf+0.83
and for the SFRTOT sample, \rstarf=\rf+0.44.
\label{lastpage}
\end{document}